\documentclass[twocolumn]{aastex631}

\usepackage{xspace, amsmath}
\usepackage{CJKutf8}

\newcommand{\as}{\arcsec\xspace}

\received{August 26, 2022}
\revised{October 19, 2022}
\accepted{\today}

\shorttitle{ALMA-MaNGA PSBs}
\shortauthors{Otter et al.}

\begin{document}
\begin{CJK*}{UTF8}{bsmi}

\title{Resolved Molecular Gas Observations of MaNGA Post-starbursts Reveal a Tumultuous Past}

\correspondingauthor{Justin Atsushi Otter}
\email{jotter2@jhu.edu}

\author[0000-0002-0786-7307]{Justin Atsushi Otter}
\affiliation{William H. Miller III Department of Physics and Astronomy, Johns Hopkins University, Baltimore, MD 21218, USA}

\author[0000-0001-7883-8434]{Kate Rowlands}
\affiliation{AURA for ESA, Space Telescope Science Institute,
3700 San Martin Drive, Baltimore, MD 21218, USA}
\affiliation{William H. Miller III Department of Physics and Astronomy, Johns Hopkins University, Baltimore, MD 21218, USA}

\author[0000-0002-4261-2326]{Katherine Alatalo}
\affiliation{Space Telescope Science Institute, 3700 San Martin Dr, Baltimore, MD 21218, USA}
\affiliation{William H. Miller III Department of Physics and Astronomy, Johns Hopkins University, Baltimore, MD 21218, USA}

\author[0000-0003-0486-5178]{Ho-Hin Leung}
\affiliation{SUPA, School of Physics \& Astronomy, University of St Andrews, North Haugh, St Andrews, Fife KY16 9SS, UK}

\author[0000-0002-8956-7024]{Vivienne Wild}
\affiliation{SUPA, School of Physics \& Astronomy, University of St Andrews, North Haugh, St Andrews, Fife KY16 9SS, UK}

\author[0000-0002-0696-6952]{Yuanze Luo}
\affiliation{William H. Miller III Department of Physics and Astronomy, Johns Hopkins University, Baltimore, MD 21218, USA}

\author[0000-0003-4030-3455]{Andreea O. Petric}
\affiliation{Space Telescope Science Institute, 3700 San Martin Dr, Baltimore, MD 21218, USA}

\author[0000-0001-6245-5121]{Elizaveta Sazonova}
\affiliation{William H. Miller III Department of Physics and Astronomy, Johns Hopkins University, Baltimore, MD 21218, USA}

\author{David V. Stark}
\affiliation{Space Telescope Science Institute, 3700 San Martin Dr, Baltimore, MD 21218, USA}
\affiliation{Department of Physics and Astronomy, Haverford College, 370 Lancaster Avenue, Haverford, PA 19041, USA}

\author[0000-0001-6670-6370]{Timothy Heckman}
\affiliation{William H. Miller III Department of Physics and Astronomy, Johns Hopkins University, Baltimore, MD 21218, USA}

\author[0000-0003-4932-9379]{Timothy A. Davis}
\affiliation{Cardiff Hub for Astrophysics Research \& Technology, School of Physics \& Astronomy, Cardiff University, Queens Buildings, The Parade, Cardiff, CF24 3AA, UK}

\author[0000-0002-1768-1899]{Sara Ellison}
\affiliation{Department of Physics and Astronomy, University of Victoria, Victoria, British Columbia V8P 1A1, Canada}

\author[0000-0002-4235-7337]{K. Decker French}
\affiliation{Department of Astronomy, University of Illinois, 1002 W. Green St., Urbana, IL 61801, USA}

\author[0000-0003-0215-1104]{William Baker}
\affiliation{Kavli Institute for Cosmology, University of Cambridge, Madingley Road, Cambridge, CB3 0HA, UK}
\affiliation{Cavendish Laboratory - Astrophysics Group, University of Cambridge, 19 JJ Thompson Avenue, Cambridge, CB3 0HE, UK}

\author[0000-0001-6395-4504]{Asa F. L. Bluck}
\affiliation{Department of Physics, Florida International University, 11200 SW 8th Street, Miami, Florida 33199, USA}

\author[0000-0002-3249-8224]{Lauranne Lanz}
\affiliation{Department of Physics, The College of New Jersey, Ewing, NJ 08618, USA}

\author[0000-0001-7218-7407]{Lihwai Lin}
\affiliation{Institute of Astronomy and Astrophysics, Academia Sinica, No. 1, Section 4, Roosevelt Road, Taipei 10617, Taiwan}

\author[0000-0002-4314-8713]{Charles Liu}
\affiliation{Department of Physics \& Astronomy, City University of New York, College of Staten Island, 2800 Victory Blvd, Staten Island, NY 10314, USA}
\affiliation{Department of Astrophysics \& Hayden Planetarium, American Museum of Natural History, Central Park West at 79th Street, New York, NY 10024, USA}
\affiliation{Physics Program, The Graduate Center, CUNY, New York, NY 10016, USA}

\author[0000-0003-1045-0702]{Carlos López Cobá}
\affiliation{Institute of Astronomy and Astrophysics, Academia Sinica, No. 1, Section 4, Roosevelt Road, Taipei 10617, Taiwan}

\author[0000-0003-0846-9578]{Karen L Masters}
\affiliation{Department of Physics and Astronomy, Haverford College, 370 Lancaster Avenue, Haverford, PA 19041, USA}

\author{Preethi Nair}
\affiliation{University of Alabama, Department of Physics and Astronomy, Tuscaloosa, AL 35487, USA}

\author[0000-0002-1370-6964]{Hsi-an Pan (潘璽安)}
\affiliation{Department of Physics, Tamkang University, No.151, Yingzhuan Road, Tamsui District, New Taipei City 251301, Taiwan }

\author[0000-0003-0483-3723]{Rogemar A. Riffel}
\affiliation{Departamento de Física, CCNE, Universidade Federal de Santa Maria, Santa Maria, RS 97105-900, Brazil}
\affiliation{Laboratório Interinstitucional de e-Astronomia - LIneA, Rua Gal. José Cristino 77, Rio de Janeiro, RJ - 20921-400, Brazil}

\author[0000-0002-8798-3972]{Jillian M. Scudder}
\affiliation{Department of Physics \& Astronomy, Oberlin College, Oberlin, OH 44074, USA}

\author[0000-0003-2599-7524]{Adam Smercina}
\affiliation{Astronomy Department, University of Washington, Seattle, WA 98195, USA}

\author[0000-0002-6301-638X]{Freeke van de Voort}
\affiliation{Cardiff Hub for Astrophysics Research \& Technology, School of Physics \& Astronomy, Cardiff University, Queens Buildings, The Parade, Cardiff, CF24 3AA, UK\\}

\author[0000-0003-1614-196X]{John R. Weaver}
\affiliation{Department of Astronomy, University of Massachusetts, Amherst, MA 01003, USA}
\affiliation{Cosmic Dawn Center (DAWN), Jagtvej 128, DK-2200 Copenhagen N, Denmark}
\affiliation{Niels Bohr Institute, University of Copenhagen, Jagtvej 128, København N, DK-2200, Denmark}

\begin{abstract}
Post-starburst galaxies (PSBs) have recently and rapidly quenched their star-formation, thus they are an important way to understand how galaxies transition from star-forming late-types to quiescent early-types. 
The recent discovery of large cold gas reservoirs in PSBs calls into question the theory that galaxies must lose their gas to become quiescent.
Optical Integral Field Spectroscopy (IFS) surveys have revealed two classes of PSBs: central PSBs with central quenching regions and ring PSBs with quenching in their outskirts.
We analyze a sample of 13 nearby ($z < 0.1$) PSBs with spatially resolved optical IFS data from the Mapping Nearby Galaxies at Apache Point Observatory (MaNGA) survey and matched resolution Atacama Large (sub-)Millimeter Array (ALMA) observations of $^{12}$CO(1-0).
Disturbed stellar kinematics in 7/13 of our PSBs and centrally concentrated molecular gas is consistent with a recent merger for most of our sample. In galaxies without merger evidence, alternate processes may funnel gas inwards and suppress star-formation, which may include outflows, stellar bars, and minor mergers or interactions.
The star-formation efficiencies of the post-starburst regions in nearly half our galaxies are suppressed while the gas fractions are consistent with star-forming galaxies.
AGN feedback may drive this stabilization, and we observe AGN-consistent emission in the centers of 5/13 galaxies.
Finally, our central and ring PSBs have similar properties except the ionized and molecular gas in central PSBs is more disturbed. Overall, the molecular gas in our PSBs tends to be compact and highly disturbed, resulting in concentrated gas reservoirs unable to form stars efficiently.
\end{abstract}

\keywords{}

\section{Introduction} \label{sec:intro}

The global star-formation rates of galaxies form a bimodal distribution, resulting in a division between star-forming late-type galaxies and quiescent early-type galaxies \citep{kauffmann_sfh_2003}. 
The galactic color-magnitude diagram quantifies this divide with primarily early-type galaxies occupying the red sequence and late-type galaxies dominating the blue cloud \citep{baldry_quantifying_2004, jin_color-magnitude_2014}.
Blue cloud galaxies are generally star-forming, gas-rich, and have disk morphologies, whereas red sequence galaxies are predominantly quiescent and gas-poor \citep{blanton_properties_2009}. 

A significant population of galaxies have transitioned from the blue cloud to the red sequence over the last $\sim$6 Gyr \citep{bell_star_2007, bell_what_2012, ilbert_mass_2013}. 
Galaxies undergoing such a transition must somehow quench their star-formation and eventually deplete their gas reservoirs.
While many quenching and gas expulsion mechanisms have been proposed, there is much work remaining to fully understand the relative importance of these processes through cosmic time.

A successful quenching mechanism must either deplete the galaxy's gas reservoirs or suppress star-formation by making it more difficult for the gas to collapse and form stars.
Galaxies could lose their molecular gas by expelling their gas through outflows \citep[e.g.][]{feruglio_quasar_2010, baron_evidence_2017}, exhausting their gas reservoirs in a starburst or from a lack of gas accretion \citep{bekki_passive_2002, davis_atlas3d_2011}, and finally through environmental processes such as ram pressure stripping \citep{gunn_infall_1972, chung_vla_2009} or fast galaxy interactions \citep{mihos_morphology_1995, moore_galaxy_1996, bekki_unequal-mass_1998}.
While these mechanisms may be important in removing gas from galaxies as they transition to the red sequence, the existence of quenched and quenching galaxies with significant molecular gas reservoirs demonstrates that quenching mechanisms that stabilize the molecular gas against star-formation are also important.
These processes suppressing star-formation could include active galactic nuclei (AGN) and stellar feedback \citep[e.g.][]{heckman_1990, kaviraj_2007, cicone_2014}, morphological quenching \citep{martig_morphological_2009}, and stellar bars \citep{salim_spinning_2020}.

The ability of the gas to collapse and form stars can be captured by the star-formation efficiency (SFE) of a galaxy, the star-formation rate per molecular gas mass.
On a cloud scale, the SFE is governed by sub-kpc gas properties such as turbulence \citep{krumholz_general_2005, murray_star_2011, federrath_star_2012, krumholz_universal_2012}, whereas global star-formation properties of galaxies are impacted by global galaxy properties and the galaxy's environment \citep{bluck_are_2020}.
Thus, to fully understand the evolution of gas and star-formation in a galaxy, both global and local scales must be considered.

To gain a better understanding of how various quenching mechanisms transform star-forming galaxies to quiescent ones, we study the population of galaxies currently undergoing this transition.
\citet{schawinski_green_2014} found that optically green colors alone are not enough to classify a transitioning population, as many late-type green valley galaxies are typical spiral galaxies with a larger buildup of intermediate-age stars or an excess of dust.
Post-starburst galaxies (PSBs) are ideal for studying short-timescale evolution to quiescence as they have recently (in the last Gyr) and rapidly quenched their star-formation. 

Classical post-starbursts, or ``E/K+A" galaxies, are galaxies with a dominant intermediate-age (A-star) stellar population and a lack of emission lines typically associated with star-formation, such as H$\alpha$ or [O II]$\lambda$3727Å.
These characteristics indicate a recent, rapid quenching of star-formation after a starburst, leaving a dominant A-type stellar population (typically identified with Balmer series absorption lines) and little-to-no on-going star-formation. 
Though classical post-starbursts have indeed quenched their star-formation recently, they represent an older population of transitioning galaxies as the strict cut on emission line strength removes post-starburst galaxies at the beginning of their transition.
Emission line cuts can select against PSBs hosting energetic processes that produce emission lines, such as shocks and AGN, though more recent E+A selection methods allow for lines associated with these processes such as [NII] \citep[e.g.][]{wild_bursty_2007, greene_refining_2021}.
Other PSB selection methods require emission line ratios consistent with shock excitation \citep{alatalo_shocked_2016}, or have no emission line requirements at all \citep{wild_bursty_2007, wild_psbs_2009, rowlands_evolution_2015, rowlands_sdss-iv_2018}.
Different post-starburst selection methods result in overlapping but distinct samples with different physical properties.
A detailed discussion of post-starburst selection criteria can be found in \citet{french_evolution_2021}.

Recently, a number of studies have begun to probe the conditions of the interstellar medium (ISM) in PSBs, yielding insights into the mechanisms driving star-formation suppression in these galaxies.
Crucially, these studies find that PSBs have significant molecular gas reservoirs \citep{french_discovery_2015, rowlands_evolution_2015, alatalo_shocked_2016-1} and dust content \citep{alatalo_twilight_2017, smercina_after_2018, li_evolution_2019} despite their lack of on-going star-formation.
Thus gas removal or exhaustion is not necessary for star-formation suppression in PSBs, but the unresolved nature of most of these studies makes it difficult to determine why the gas is not forming stars.

The optical integral field unit (IFU) survey `Mapping Nearby Galaxies at Apache Point Observatory' (MaNGA) has revealed spatially resolved post-starburst regions within galaxies \citep{chen_post-starburst_2019, greene_refining_2021}. 
\citet{chen_post-starburst_2019} differentiate between central and ring post-starbursts depending on where the post-starburst regions reside.
Central PSBs (cPSBs) include galaxies with central regions dominated by post-starburst signatures and show evidence of a recent disruptive event such as a major merger, likely leading to global quenching.
Ring PSBs (rPSBs) show post-starburst signatures on the outskirts of the galaxy - though not necessarily forming a full `ring', with star-formation on-going in the center, potentially due to a disruption of gas fuelling to the outer regions.
\citet{chen_post-starburst_2019} show that cPSBs and rPSBs have different kinematic properties and star-formation histories, indicating that these two types of PSBs have distinct origins and are not sequential evolutionary phases.
Further, the overall evolution of rPSBs is uncertain as it is unclear whether an rPSB phase leads to global galaxy quenching or if this is a temporary phase and star-formation is re-invigorated in the outskirts.

We analyze a sample of 13 nearby ($z < 0.1$) post-starburst galaxies with MaNGA and Atacama Large (sub-)Millimeter Array (ALMA) data to characterize the molecular gas and star-formation properties on kpc scales in PSBs.
We adopt a flat $\Lambda$CDM cosmology with $H_0$ = 70 km/s/Mpc, $\Omega_m$ = 0.30, and $\Omega_\Lambda$ = 0.70. 
This paper is organized as follows. Section~\ref{sec:obs} presents the MaNGA and ALMA data and our sample of post-starburst galaxies.
Section~\ref{sec:methods} describes our methods of computing molecular gas masses from the ALMA data and the spectral fitting of the MaNGA IFU data to obtain galaxy properties such as star-formation rates.
We present our results in Section~\ref{sec:results} and discuss the implications of our results in Section~\ref{sec:discuss}.

\section{Observations and Sample} \label{sec:obs}

\subsection{The MaNGA Survey}

The MaNGA survey \citep{bundy_overview_2014} consists of optical IFU observations for 10,000 galaxies with $0.01 < z < 0.15$, taken with the Baryon Oscillation Spectroscopic Survey (BOSS) spectrograph \citep{smee_multi-object_2013, drory_manga_2015} on the 2.5m Sloan Digital Sky Survey (SDSS) telescope \citep{gunn_25_2006} as one of three major SDSS-IV programs \citep{blanton_sloan_2017}.
Each galaxy is covered by 19-127 hexagonal fiber bundles, covering a 12-32\as diameter on the sky.
The resulting data cubes have a point-spread function with a full width at half-maximum (FWHM) of $\sim$2.5\arcsec, and have wavelength coverage from 3600–10300\AA\ with $R\sim$2000.
MaNGA observations are calibrated with standard stars observed simultaneously, with an absolute calibration of better than 5\% for $\sim$90\% of the wavelength range \citep{yan_sdss-ivmanga_2015}.
Two thirds of MaNGA galaxies belong to the Primary sample with fields of view covering 1.5 times the effective radius ($R_e$), while the remaining third, the Secondary sample, have coverage out to 2.5 $R_e$.
Both samples are selected with a flat stellar mass distribution and have mean physical resolutions of 1.37 and 2.5 kpc for the Primary and Secondary samples respectively.
More information on the MaNGA observing strategy, calibration, and survey design can be found in \citet{law_observing_2015, yan_sdss-ivmanga_2015} and \citet{wake_sdss-iv_2017}.

Throughout this work, we use MaNGA emission line properties and stellar velocities from the MaNGA Data Analysis Pipeline \citep[DAP,][]{westfall_dap_2019, belfiore_dap_2019}, accessed through \texttt{Marvin} \citep{marvin_paper_2019}.

\subsection{Sample selection} \label{sec:sample}

We first selected PSBs from the MaNGA survey in the SDSS data release 15 (DR15), including approximately half of the galaxies in the completed survey of 10,000 galaxies \citep{aguado_fifteenth_2019}.
We started with the 68 galaxies in the central and ring PSB sample of \citet{chen_post-starburst_2019} (excluding the irregular PSBs) with additional PSBs from the principal component analysis (PCA) selection method of \citet{rowlands_sdss-iv_2018}, adding another 25 galaxies, for a total of 93 galaxies.
\citet{chen_post-starburst_2019} select post-starburst spaxels with H$\delta_A$ $>$ 3\AA, W(H$\alpha$) $<$ 10\AA, and log W(H$\alpha$) $<$ 0.23 $\times$ H$\delta_A-0.46$, where H$\delta_A$ is the Lick index measuring the amount of H$\delta$ absorption and W(H$\alpha$) is the equivalent width of H$\alpha$ in emission.

PSBs are selected with the PCA method by requiring that the central 0.5$R_e$ (the $r$-band half-light radius) is visually dominated by post-starburst spaxels.
This threshold was chosen to capture PSBs selected with other methods, such as the central 3\arcsec\, fiber from SDSS Data Release 7 in \citet{wild_bursty_2007}, or spatially resolved studies like \citet{chen_post-starburst_2019}.
With this criterion, we find that 44/4706 DR15 galaxies are PSBs, similar to the 1-3\% prevalence of PSBs in the local universe \citep[e.g.][]{goto_2003, rowlands_sdss-iv_2018}.
The ``eigenspectra", or principal components, are from \citet{wild_bursty_2007}, and were generated from model galaxies with star-formation histories containing starbursts of various burst mass fractions, durations, and burst times, with simple stellar population templates from \citet{bruzual_stellar_2003}.
The first principal component (PC1) corresponds to the 4000\AA\, break strength (the $D_n4000$ index), and the second (PC2) is the excess Balmer absorption over what is expected given the 4000\AA\, break strength.
The MaNGA spectrum for each spaxel with stellar velocity uncertainty $<$ 500 km/s is shifted to rest-frame and projected onto the model-derived principal components to determine the contribution of each.
Post-starburst spaxels are identified primarily by their strong Balmer absorption (PC2), as in \citet{rowlands_sdss-iv_2018}.
After the sample selection, we slightly alter the classification boundaries, which we discuss in more detail in Appendix~\ref{app:pca}. 

Of the 93 PSBs from the combined \citet{chen_post-starburst_2019} sample and PCA sample, 23 have Dec. $< 27^\circ$ and are thus observable by ALMA.
To estimate the required sensitivity for our CO(1-0) ALMA observations, we used the integrated fluxes from single dish CO(1-0) observations when available.
Otherwise, we estimated the gas fraction and molecular gas mass of each galaxy with the mid-infrared Wide-field Infrared Survey Explorer (WISE) colors following \citet{yesuf_molecular_2017}, or conservatively assumed a gas fraction of 3\% when WISE data are not available.
We selected 3\% as this is a reasonable constraining limit on the gas fraction for PSBs \citep[e.g.][]{french_clocking_2018}.
To limit the integration time, we selected 13 galaxies with required sensitivities $>4$ mJy/beam for a 5$\sigma$ CO(1-0) detection and without previous CO observations of similar kpc scale resolution.

Our spectral fitting results (discussed in Section~\ref{sec:fitting}) reveal that one galaxy we observe with ALMA, 8939-3703, is a star-forming interloper, despite deep Balmer absorption.
We therefore do not consider this galaxy in our main analysis, though we present the ALMA and MaNGA data for this galaxy in Appendix~\ref{app:8939}. 

One rPSB in MaNGA, 8955-3701, has similar ALMA CO(1-0) observations as part of the ALMA-MaNGA QUEnching and STar-formation (ALMaQUEST) survey\footnote{\url{arc.phys.uvic.ca/~almaquest/}} \citep{lin_almaquest_2020}.
We include this galaxy in our final sample for a total of 13 galaxies with 9 cPSBs and 4 rPSBs.

Two galaxies in our final sample are not in the \citet{chen_post-starburst_2019} sample and are selected from the PCA method, 8086-3704, 9088-9102, and the star-forming interloper 8939-3703.
A key advantage of the PCA selection is that PC2 includes the excess absorption of all Balmer lines simultaneously, and thus can be applied to spaxels of a much lower S/N than traditional Balmer absorption measurements, such as H$\delta$ absorption. 
Additionally, emission lines are masked, so our sample does not exclude PSBs with emission lines excited by AGN \citep{yan_origin_2006, yan_deep2_2009, wild_bursty_2007}, shocks \citep{alatalo_shocked_2016}, and younger post-starbursts which have not totally ceased star-formation.

\begin{figure*}
    \centering
    \includegraphics[width=0.97\textwidth]{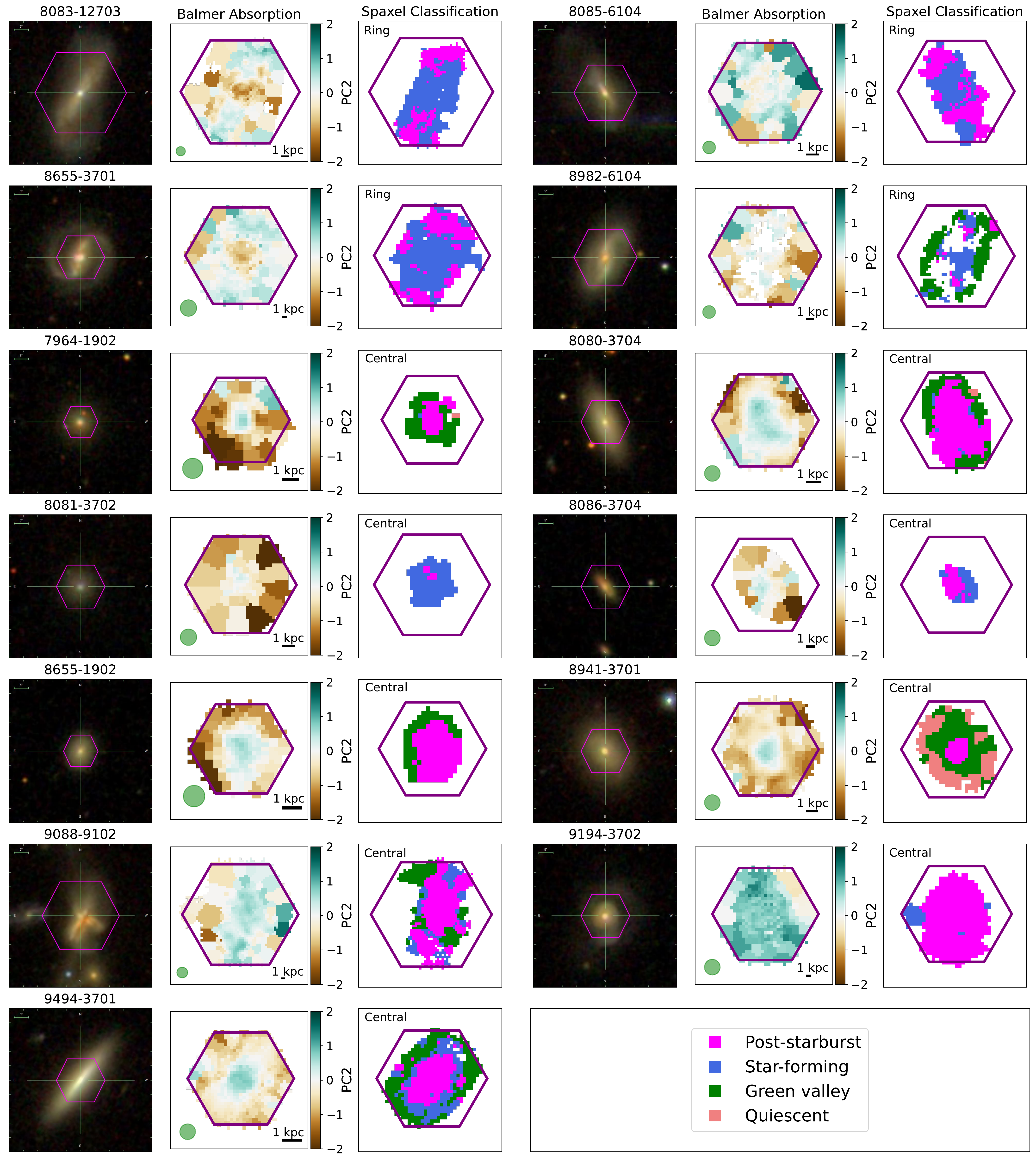}
    \caption{PCA classification results for our sample.
    Three images show the SDSS 3-color image for each galaxy, the excess Balmer absorption (PC2), and the PCA classification map.
    The MaNGA Plate-IFU number is above the first panel.
    The excess Balmer absorption is measured with the PCA method from \citet{rowlands_sdss-iv_2018}, where positive values (in teal) indicate post-starburst regions.
    The length of the black rectangle in the lower right shows the spatial scale of 1 kpc.
    The green ellipse in the bottom left shows the 2.5\arcsec\, MaNGA average seeing.
    Finally, in the PCA classification map post-starburst spaxels are magenta, star-forming spaxels are blue, green valley spaxels are green, and quiescent spaxels are light red.
    The top left text states whether the galaxy is classified as a central or ring PSB.
    The purple hexagons show the MaNGA IFU coverage.}
    \label{fig:overview_pca}
\end{figure*}

Figure~\ref{fig:overview_pca} shows the SDSS 3-color image, the PCA spaxel classification maps, and excess Balmer absorption for each galaxy in our sample.
We compare our spaxel classification maps to the ring and central PSB classifications made by \citet{chen_post-starburst_2019}, and find consistent determinations by visual inspection with the exceptions of 8080-3704 and 9194-3702, where the PCA classification shows a central post-starburst morphology, but they are classified as rPSBs in \citet{chen_post-starburst_2019}.
We classify these galaxies as cPSBs for the remainder of our analysis for consistency with our PCA classification maps.
Additionally, 8982-6104 is a ring PSB but the PCA classification was unable to accurately decompose the spectra of a number of central spaxels, potentially due to AGN contamination, or a strong starburst.
However, we still identify this as an rPSB, even though the ring quenching region is patchy.
We stress that there are a plethora of different post-starburst selection criteria, and no single selection will include all spectra considered as `post-starburst' \citep{french_evolution_2021}.

\begin{figure*}
    \centering
    \includegraphics[width=0.9\textwidth]{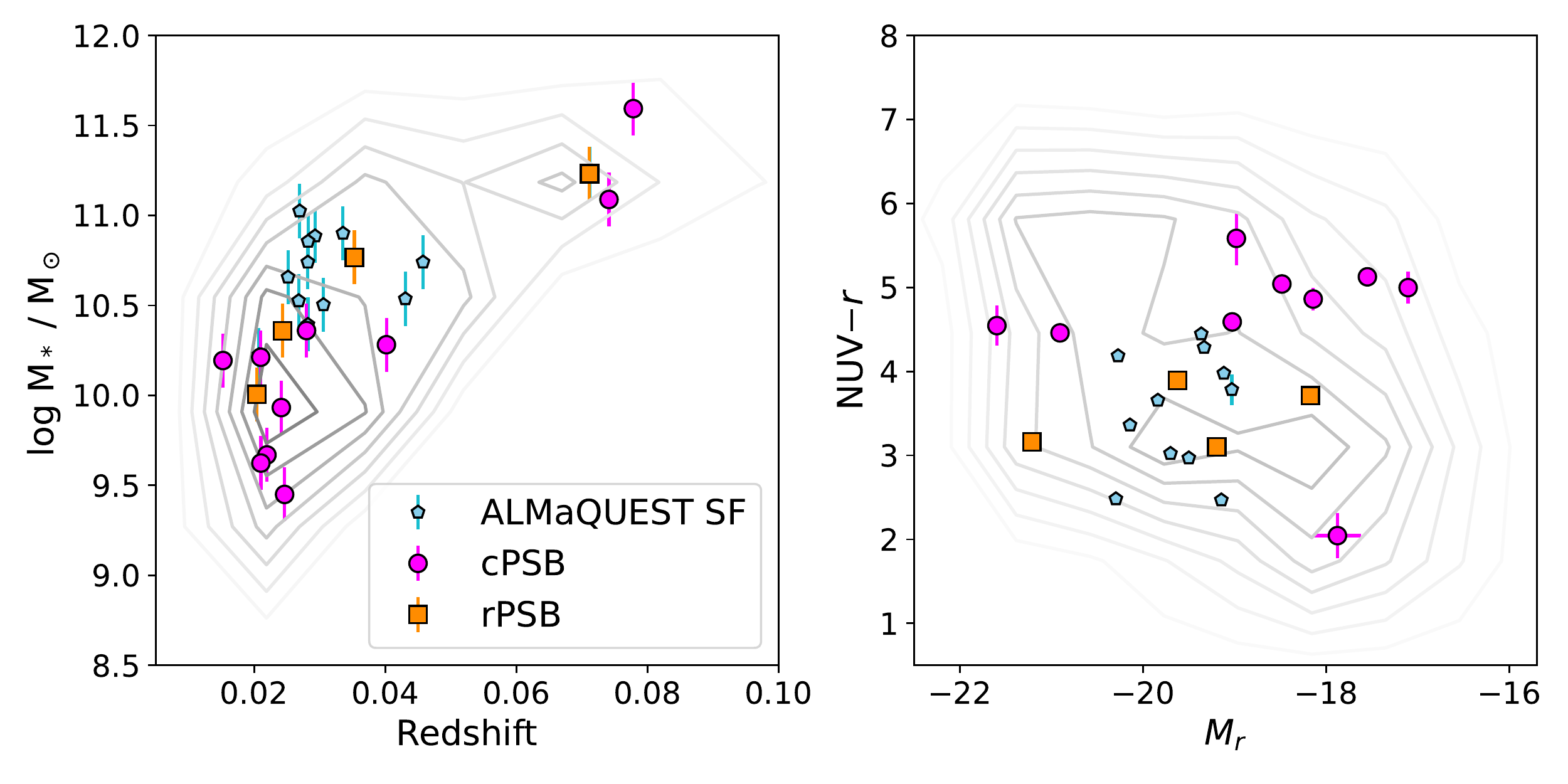}
    \caption{Left: stellar mass and redshift diagram of our sample of central and ring PSBs, ALMaQUEST star-forming galaxies, and the parent MaNGA DR15 sample as gray contours. We use stellar masses from \citet{pace_resolved_2019}, and see that the majority of our PSB sample have lower masses than the ALMaQUEST star-forming sample. Right: Color-magnitude diagram with the same galaxies plotted. Magnitudes are from the NASA-Sloan Atlas. Errorbars are present for all points but may be smaller than the marker size. Our cPSBs tend to be redder than the star-forming sample while our rPSBs have similar colors.}
    \label{fig:context}
\end{figure*}

We plot the stellar mass versus redshift and color magnitude diagram of our final sample of 13 galaxies and the parent MaNGA DR15 sample in contours in Figure~\ref{fig:context}.
Magnitudes and redshifts are obtained from the NASA-Sloan Atlas\footnote{\url{http://nsatlas.org/}} (NSA), and stellar masses from \citet{pace_resolved_2019} with a color mass-to-light relation aperture correction. 
We use an uncertainty of 0.15 dex for the integrated stellar masses based on uncertainty estimates from \citet{pace_resolved_2019}.
We also plot the star-forming ``main-sequence" sample of 12 galaxies from ALMaQUEST, which have a specific star-formation rate (sSFR) $>$ $10^{-10.5}$ yr$^{-1}$ \citep{lin_almaquest_2022}.
The main-sequence sample (hereafter star-forming) also excludes the 12 ALMaQUEST galaxies selected specifically for their starburst properties in \citet{ellison_starburst_2020}, making up the ALMaQUEST starburst sample.
The remaining 22 ALMaQUEST galaxies constitute the `green valley' sample with low sSFRs indicating on-going quenching.
The main sequence galaxies are a convenient comparison sample for our PSBs because of their similar ALMA observations, though we emphasize that the majority of our sample have significantly lower stellar masses.

We also note that the spectral fitted stellar masses from \citet{pace_resolved_2019} for our post-starbursts are not fully reliable due to the unique and complex star-formation histories of PSBs.
Hence, we include these total stellar masses not as precise values but as a rough estimate for our sample.

In the color-magnitude diagram shown in Figure~\ref{fig:context}, our cPSBs tend to be redder than the ALMaQUEST star-forming galaxies as we expect for a globally quenching population.
Our rPSBs have comparable colors to the star-forming galaxies, as is consistent with their actively star-forming centers.

\subsection{ALMA observations}
We obtain ALMA 12m array band 3 observations for 13 galaxies (12 PSBs and 1 interloper) in ALMA proposal 2019.1.01136.S (P.I. Rowlands), listed in Table~\ref{tab:obs}.
Each observation has four 1.875 GHz wide spectral windows, one covering the redshifted $^{12}$CO(1-0) line with a rest frequency of 115.27 GHz, and the other three covering nearby redshifted CH$_3$OH, CN, and HC$_3$N lines with rest frequencies of 107.02 GHz, 113.5 GHz, and 100.07 GHz respectively.
These observations have minimum baselines of 15 m and a maximum of 314 m (corresponding to configuration C-2) with the exception of 9088-9102, which had a maximum baseline of 440 m.
Integration times range from 300 - 6400 s.
The native velocity resolution is 3 km/s and the native angular resolutions range from 2.5--3\arcsec.

The observations were pipeline calibrated and imaged with the Common Astronomy Software Applications package \citep[CASA (ver. 6.4.3),][]{mcmullin_casa_2007}.
We construct data cubes from each spectral window and image the continuum with the \texttt{tclean} task.
In each case, we use a circular clean mask centered on the middle of the image (approximately co-incident with the optical center of each galaxy) and with a radius extending slightly beyond the extent of the CO emission.
Following the ALMaQUEST survey \citep{lin_almaquest_2020}, we use a circular restoring beam with a FWHM of 2.5\arcsec\, and specify the pixel size of our millimeter data as 0.5\arcsec\, to match the seeing and pixel scale of the MaNGA data for an optimal comparison between the ALMA and MaNGA data.
We use Briggs weighting with a robust parameter of 0.5 and a flux threshold of the 1$\sigma$ noise level for all data products except the non-CO spectral windows where we use a robust parameter of 2 to maximize sensitivity.

We image the continuum using all four of the spectral windows, flagging out channels from the CO spectral window within 1000 km/s of the redshifted CO(1-0) frequency.
We include the entirety of the non-CO spectral windows because we do not detect any non-CO line emission.

We apply continuum subtraction to the galaxies with continuum detections with the CASA task \texttt{uvcontsub}.
We image the CO spectral windows with a spectral channel width of 11 km/s.
For direct comparison between the CO and MaNGA data, we employ the \texttt{imregrid} task in CASA to match the pixel locations in both images.

\begin{figure*}
    \centering
    \includegraphics[width=\textwidth]{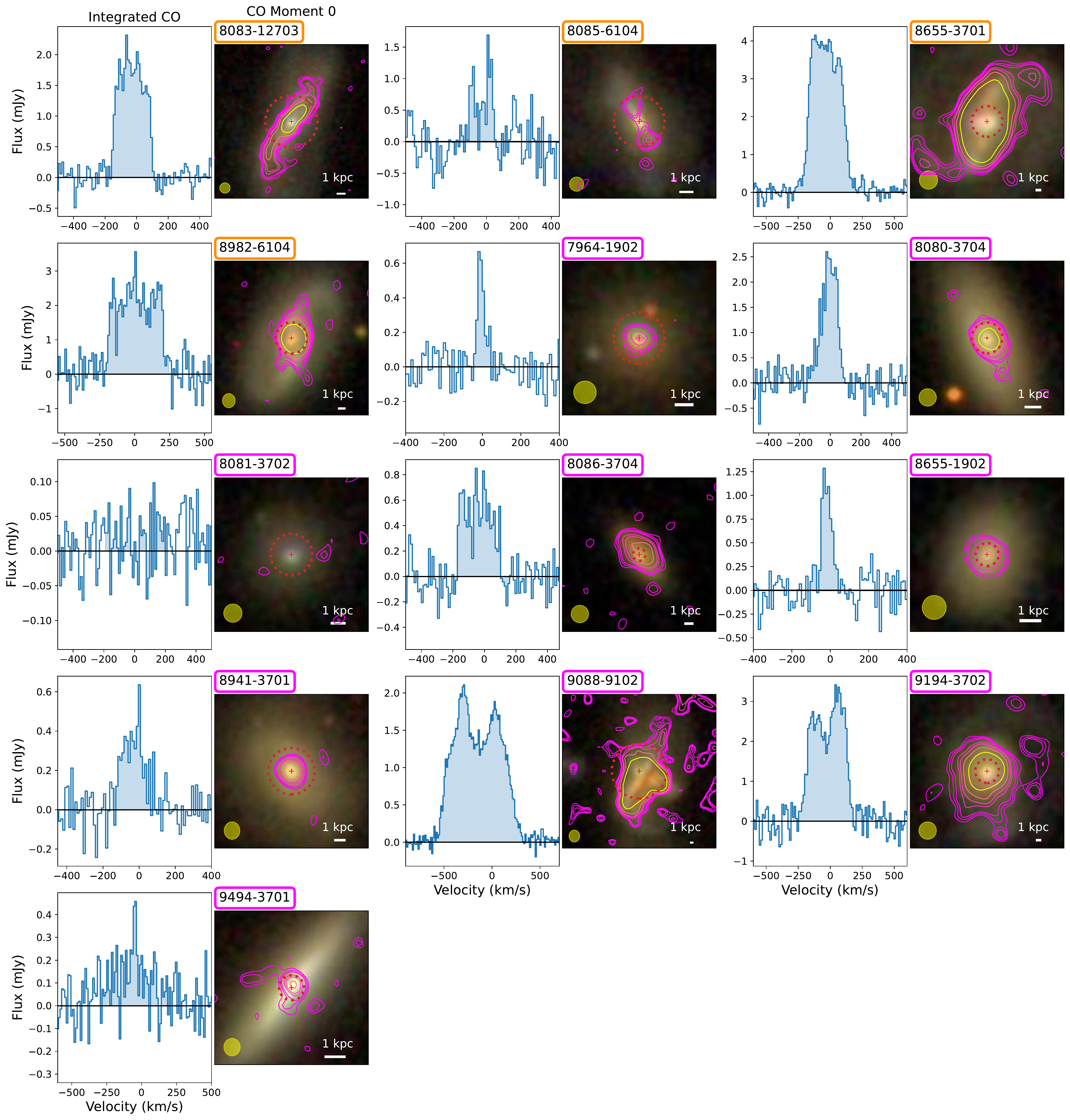}
    \caption{CO(1-0) spectra and integrated intensity contours for the 13 PSBs in our sample.
    The spectra is spatially integrated over 1.5$R_{e}$.
    The shaded region shows the spectral integration bounds for the intensity maps.
    The MaNGA Plate-IFU identifier is boxed in purple for central PSBs and orange for ring PSBs.
    The CO moment 0 contours correspond to S/N values of 3, 4, 5, 8, 10, 15, and 30, and are plotted over the SDSS 3-color image.
    The 1$\sigma$ noise levels for each galaxy are given in Table~\ref{tab:obs}.
    The red plus is the optical center of the galaxy, and the red dashed line is the $r$-band effective radius from the NSA.
    The yellow circle in the lower left is the ALMA beam size.}
    \label{fig:overview_co}
\end{figure*}

We generate integrated CO(1-0) intensity, velocity, and velocity dispersion maps for each galaxy (the moment 0, 1, and 2 maps respectively).
We set the velocity bounds of the moment maps based on the width of the CO(1-0) line in the total spectrum within 1.5 $R_e$ of each galaxy, as shown in Figure~\ref{fig:overview_co}.
We use the $r$-band half-light radius from the NSA as $R_e$ throughout this work.
For the integrated intensity, we do not apply any sigma clipping, while we remove pixels less than the 4$\sigma$ noise level for the velocity and velocity dispersion maps. 

We image the remaining three continuum subtracted spectral windows, centered on CH$_3$OH, CN, and HC$_3$N lines, with a Briggs robust parameter of 2 in order to maximize the sensitivity of our data, and a similar flux threshold of 1$\sigma$ and spectral channel width of 11 km/s, using the native beam size.
We do not detect these emission lines in any of our galaxies.

Both the imaging and moment map creation follow the methods of the ALMaQUEST survey \citep{lin_almaquest_2020}, including the use of the MaNGA beam size, allowing for a robust comparison of our sample with their similar observations of star-forming and green valley galaxies.
As mentioned in Section~\ref{sec:sample}, we include one ring PSB from the ALMaQUEST survey in our sample, 8655-3701.
The data for this galaxy is slightly different from the rest of our sample with a lower spectral resolution CO spectral window and a single dedicated continuum spectral window.
We image the continuum as above, though with only the single continuum spectral window and the channels of the CO spectral window at least 1000 km/s away from the expected CO line frequency.
We image the CO spectral window in an identical way to the rest of our sample, as our chosen channel width of 11 km/s is equivalent to the spectral resolution of this data.
We do not have spectral coverage of the CH$_3$OH, CN, and HC$_3$N lines for this galaxy.

As shown in Table~\ref{tab:obs}, the native beam sizes of our observations tend to be slightly higher than the MaNGA seeing of 2.5\arcsec.
To ensure that our choice of a smaller restoring beam does not impact our results, we image our data products with the native beam size and find similar results with no systematic offsets.
The data cubes with the selected 2.5\arcsec\, restoring beam tend tend to have slightly higher noise levels than in the data cubes imaged with the native restoring beam, up to $\sim$10\% for the cubes with the largest native beam sizes.
For consistency, we use the 2.5\arcsec\, restoring beam fluxes and noise maps, which may yield a slight overestimate of our uncertainties.

\begin{table*}
    \centering
    \begin{tabular}{p{2.8cm}|p{3cm}|p{2.5cm}|p{2.3cm}|p{2.5cm}}
    MaNGA Plate-IFU (1) & Observation time (s) (2)& Observation Date (3) & Line sensitivity (mJy/beam) (4) & Native beam size ($\theta_{maj}\arcsec \times \theta_{min} \arcsec$) (5) \\ \hline
    7964-1902 & 695.5 & Nov 24, 2019 & 0.9 & $2.8\arcsec \times 2.3\arcsec$ \\
    8080-3704 & 302.4 & Dec 23, 2019 & 1.1 & $3.1\arcsec \times 2.5\arcsec$ \\
    8081-3702 & 5745.6 & Dec 25, 2019 & 0.3 & $2.9\arcsec \times 2.3\arcsec$ \\
    8083-12703 & 302.4 & Dec 14, 2019 & 0.7 & $2.9\arcsec \times 2.2\arcsec$ \\
    8085-6104 & 302.4 & Dec 14, 2019 & 1.5 & $2.8\arcsec \times 2.3\arcsec$ \\
    8086-3704$^*$ & 302.4 & Dec 14, 2019& 1.1 & $2.7\arcsec \times 2.4\arcsec$ \\
    8655-1902 & 302.4 & Dec 13, 2019 & 1.7 & $3.1\arcsec \times 2.4\arcsec$ \\
    8939-3703$^*$ & 635.0 & Dec 31, 2019 & 0.9 & $3.5\arcsec \times 2.4\arcsec$ \\
    8941-3701 & 1844.6 & Dec 14, 2019 & 0.7 & $3.3\arcsec \times 2.4\arcsec$ \\
    8982-6104 & 302.4 & Nov 23, 2019 & 1.2 & $3.5\arcsec \times 2.5\arcsec$ \\
    9088-9102$^*$ & 6531.8 & Jan 2, 2020 & 0.4 & $2.6\arcsec \times 2.5\arcsec$ \\
    9194-3702 & 302.4 & Dec 24, 2019 & 0.9 & $3.0\arcsec \times 2.5\arcsec$ \\
    9494-3701 & 3870.7 & Jan 1, 2020 & 0.6 & $2.9\arcsec \times 2.4\arcsec$ \\
    \end{tabular}
    \caption{Observations for each target from ALMA proposal 2019.1.01136.S. (1) MaNGA Plate-IFU identifier. Galaxies marked with $^*$ were selected with the PCA method and are not in the \citet{chen_post-starburst_2019} sample. (2) Integration time of ALMA observation. (3) Date of ALMA observation. (4) Computed CO line sensitivity. The line sensitivity is calculated from a line-free region of each data cube. (5) The native beam size of the observations.}
    \label{tab:obs}
\end{table*}

\section{Methods} \label{sec:methods}

\subsection{Molecular gas masses}

We compute molecular gas masses from the integrated CO(1-0) intensity maps with Equation~\ref{eq:alphaco}, following \citet{bolatto_co--h2_2013} with a constant, Milky Way conversion factor of  $\alpha_{CO} = 4.35$ M$_\odot$ (K km s$^{-1}$ pc$^2$)$^{-1}$, including a Helium correction.
\begin{equation}\label{eq:alphaco}
    M_{H_2} = \alpha_{CO} * L_{CO}
\end{equation}
The value of $\alpha_{CO}$ represents a significant source of uncertainty, especially given that $\alpha_{CO}$ can be as low as 0.8 M$_\odot$ (K km s$^{-1}$ pc$^2$)$^{-1}$ in extreme environments such as in starburst galaxies \citep{narayanan_co-h2_2011}.
We discuss our results with a variable $\alpha_{CO}$ in Appendix~\ref{app:alphaCO}.

We report global CO and continuum properties of our sample in Table~\ref{tab:global}.
More details on the continuum emission and radio detections of our sample are in Appendix~\ref{app:radio}.

\begin{table*}
\centering
\begin{tabular}{p{1.6cm}|p{2.0cm}|p{2.1cm}|p{0.8cm}|p{0.8cm}|p{1.5cm}|p{2cm}|p{0.6cm}|p{1.9cm}|p{0.8cm}}
Plate-IFU (1) & R.A. (2) & Decl. (3) & $z$ (4) & PSB type (5) & $S_{CO}$ (6) & log $M_{H_2}/M_\odot$ (7) & CO $R_{50}$ (8) & $S_{3mm}$ (9) & $\sigma_{3mm}$ (10) \\ 
 &  &  &  &  & $\mathrm{Jy\,km\,s^{-1}}$ &  & $\mathrm{kpc}$ & $\mathrm{mJy}$ & $\mathrm{mJy}$ \\ \hline
7964-1902 & $21:09:41.43$ & $0:37:39.97$ & 0.024 & cPSB & $0.8\pm0.2$ & $7.99\pm0.08$ & 0.7 & - & - \\
8080-3704 & $3:17:49.79$ & $-0:33:16.77$ & 0.021 & cPSB & $8.4\pm0.4$ & $8.86\pm0.02$ & 0.6 & $0.23\pm0.02$ & 0.035 \\
8081-3702 & $3:19:47.24$ & $0:37:25.76$ & 0.025 & cPSB & $<0.2$ & $<7$ & - & - & - \\
8083-12703 & $3:19:43.04$ & $0:33:55.72$ & 0.024 & rPSB & $11.1\pm0.2$ & $9.113\pm0.01$ & 1.7 & $0.183\pm0.005$ & 0.013 \\
8085-6104 & $3:26:50.14$ & $0:11:54.92$ & 0.020 & rPSB & $3.1\pm0.5$ & $8.41\pm0.06$ & - & - & - \\
8086-3704 & $3:48:41.67$ & $-0:39:03.60$ & 0.040 & cPSB & $3.7\pm0.2$ & $9.08\pm0.03$ & 1.6 & - & - \\
8655-1902 & $23:53:52.52$ & $-0:05:55.43$ & 0.022 & cPSB & $2.2\pm0.2$ & $8.33\pm0.04$ & 0.7 & - & - \\
8655-3701 & $23:47:00.44$ & $-0:26:50.59$ & 0.071 & rPSB & $32.1\pm0.2$ & $10.518\pm0.003$ & 3.0 & $0.819\pm0.009$ & 0.025 \\
8939-3703$^\dag$ & $8:23:37.68$ & $25:01:12.42$ & 0.021 & - & $<0.7$ & $<8$ & - & - & - \\
8941-3701 & $8:00:14.30$ & $26:41:52.85$ & 0.028 & cPSB & $1.7\pm0.1$ & $8.41\pm0.03$ & 0.8 & - & - \\
8982-6104 & $13:32:13.70$ & $26:56:59.93$ & 0.035 & rPSB & $24.5\pm0.6$ & $9.78\pm0.01$ & 1.2 & $0.53\pm0.02$ & 0.045 \\
9088-9102 & $16:09:53.36$ & $26:37:33.21$ & 0.078 & cPSB & $29.9\pm0.1$ & $10.567\pm0.002$ & 3.9 & $0.318\pm0.003$ & 0.012 \\
9194-3702 & $3:08:07.07$ & $0:27:22.35$ & 0.074 & cPSB & $24.7\pm0.5$ & $10.44\pm0.008$ & 3.0 & $0.16\pm0.02$ & 0.032 \\
9494-3701 & $8:27:01.41$ & $21:42:24.31$ & 0.015 & cPSB & $1.6\pm0.1$ & $7.88\pm0.03$ & 1.1 & $0.245\pm0.006$ & 0.012 \\
\end{tabular}
\caption{Integrated CO and continuum properties of our sample.
(1) MaNGA Plate-IFU identifier. 8939-3703, the star-forming interloper, is denoted with $^\dag$ (see Appendix~\ref{app:8939}).
(2) Optical R.A. (J2000).
(3) Optical declination (J2000). 
(4) Optical redshift.
(5) Ring (rPSB) or central (cPSB) classification.
(6) Integrated CO flux.
(7) Total molecular gas mass.
(8) CO emission half-light radius (computed with \texttt{statmorph}).
(9) 3 mm continuum flux.
(10) 1$\sigma$ noise level of 3 mm emission.}
\label{tab:global}
\end{table*}

\subsection{Spectral fitting} \label{sec:fitting}
To obtain star-formation histories, we fit the optical MaNGA spectra of the PSBs using \textsc{Bagpipes} \citep{bagpipes2018,bagpipes2019}, a fully Bayesian spectral fitting code.
We stack all spaxels classified as post-starburst through PCA (Section~\ref{sec:sample}) after careful removal of flagged spaxels in MaNGA suffering from foreground stars, dead fiber, or low signal-to-noise ratio.
Stacking is performed by simple unweighted summing, and uncertainties are summed in quadrature.
We do not correct for the stellar or gas velocities in individual spaxels.
To match the spectral sampling rate $R\sim2000$ of MaNGA spectra, we limit our fitted spectral range to rest-frame $\lambda<7500\text{\AA}$, the limit of the stellar templates from the MILES library \citep{MILES}.

\textsc{Bagpipes} uses the 2016 version of the \cite{bruzual2003} spectral synthesis models, and assumes the initial mass function from \cite{kroupa2001}.
We adopt the two-component star-formation history (SFH) functional form from \cite{wild_star_2020} designed to describe PSBs:

\begin{align}\label{eq:psb_wild2020}
    \nonumber
    \mathrm{SFR}(t) & \propto
    \frac{1-f_{\mathrm{burst}}}{\int \psi_e dt} \times \psi_e(t)\Big|_{t_{\mathrm{form}}>t>t_{\mathrm{burst}}} \\ 
    & + \frac{f_{\mathrm{burst}}}{\int \psi_{\mathrm{burst}} dt} \times \psi_{\mathrm{burst}}(t)
\end{align}

where $t_{\mathrm{form}}$ is the lookback time when the older population began to form, $t_{\mathrm{burst}}$ is the time since the peak of the starburst, and $f_{\mathrm{burst}}$ is the portion of mass formed during the starburst.
$\psi_e$ and $\psi_{\mathrm{burst}}$ are given by:

\begin{align}
    \psi_e(t) &= \exp^{\frac{-t}{\tau_e}} \\
\label{eq:dpl}
    \psi_{\mathrm{burst}}(t) &= \Big[\big(\frac{t}{t'_{\mathrm{burst}}}\big)^\alpha 
    + \big(\frac{t}{t'_{\mathrm{burst}}}\big)^{-\beta}\Big]^{-1}\;.
\end{align}

$\psi_e$ is the older, exponential decay component and $\psi_{\mathrm{burst}}$ is the double power-law starburst component.
$\tau_e$ is the older population's exponential decay timescale.
$\alpha$ and $\beta$ are the declining and increasing timescales of the burst respectively, with larger values corresponding to steeper SFH slopes.
$t'_{\mathrm{burst}}$ is the age of the universe at the peak of the starburst. 
All times in the above equations are in the age of the universe (i.e. $t$ = 13.8 Gyr is the present).

We implement a two-step metallicity evolution model, which allows for the stellar metallicity of newly formed stars before ($Z_{\rm{old}}$) and after the peak of the starburst ($Z_{\rm{burst}}$) to have independent metallicity levels as motivated by hydrodynamic simulations (Leung et al. in prep.). 
For dust attenuation, we use the two-component dust law from \cite{wild_bursty_2007} with a fixed power-law exponent $n=0.7$ for the interstellar medium and a steeper power-law exponent for stars younger than 10 Myr of $n=1.3$.
These young stars are more attenuated than the older ones by a factor $\eta$ \citep[$=1/\mu$ in][see Equation 3]{wild_bursty_2007}, as they are assumed to be surrounded by their birth clouds.
Nebular emission lines and residuals of skyline subtraction \citep[drawing from the catalogue of][]{skylines} are masked during fitting.

We include a Gaussian Process (GP) kernel as an additive model component to help account for correlated systematic errors caused by observational and calibrational problems in the observed frame as well as model-data mismatch from model limitations (stellar templates, assumed SFH etc.) in the rest-frame \citep[see Section 4.3 of][]{bagpipes2019}.
The kernel employed is a stochastically-driven damped simple harmonic oscillator (SHOTerm), implemented through the \texttt{celerite2} python package \citep{celerite1,celerite2}.
The final model has 18 parameters, with 15 free to vary with priors.
The parameter priors are listed in Appendix~\ref{app:SFH}.
Sampling of the posterior surface is performed with the nested sampling algorithm \textsc{MultiNest} \citep{multinest} and its python interface \citep{pymultinest}.
For detailed implementation of all elements of the model, its testing and potential limitations, see Leung et al. (in prep.).

Due to the recent starbursts dominating the contribution in the observed spectra as a result of their much lower mass-to-light ratio \citep[see Figure 7 of ][]{french_clocking_2018}, the stellar mass formed during the older, pre-burst component can be underestimated by a small amount, which leads to a slight overestimation in burst mass fraction ($f_\mathrm{burst}$).
However, since this affects all PSBs, it is expected to have little qualitative effects on our results (discussed in Leung et al. in prep.).

After fitting, star-formation rates (SFR) are measured as the average across the most recent $10^8$ years from the fitted SFHs.

We report the CO properties and spectral fitting results of the PSB regions for each galaxy in our sample in Table~\ref{tab:region}.
All galaxies but one aforementioned interloper (8939-3703) exhibit clear histories of recent starburst and subsequent quenching from from their fitted SFHs, as displayed in Appendix~\ref{app:SFH}.

\begin{table*}
\centering
\begin{tabular}{p{1.6cm}|p{1.6cm}|p{1.9cm}|p{1.7cm}|p{1.7cm}|p{1.5cm}|p{1.5cm}|p{1.5cm}|p{1.2cm}}
Plate-IFU (1) & $S_{CO, PSB}$ (2) & log $M_{H_2, PSB}$ (3) & log SFR (4) & log $M_{*, PSB}$ (5) & $t_\mathrm{burst}$ (6) & $f_{\mathrm{burst}}$ (7) & $F_{1Gyr}$ (8) & $A_{PSB}$ (9) \\
 & $\mathrm{Jy\,km\,s^{-1}}$ & $\log M_\odot$ & $\log M_\odot$ yr$^{-1}$ & log $M_\odot$  & $\mathrm{Gyr}$ &  &  & $\mathrm{kpc^{2}}$ \\ \hline
7964-1902 & $0.44\pm0.04$ & $7.71\pm0.04$ & $-1.7^{+0.1}_{-0.1}$ & $9.24^{+0.05}_{-0.05}$ & $1.9^{+0.3}_{-0.2}$ & $0.6^{+0.2}_{-0.2}$ & $0.049^{+0.007}_{-0.006}$ & 2.7 \\
8080-3704 & $8.0\pm0.2$ & $8.84\pm0.01$ & $-1.6^{+0.4}_{-1.0}$ & $10.06^{+0.06}_{-0.06}$ & $0.59^{+0.09}_{-0.04}$ & $0.12^{+0.02}_{-0.02}$ & $0.14^{+0.02}_{-0.02}$ & 26.0 \\
8081-3702 & $<0.01$ & $<6.0$ & $-2.4^{+0.1}_{-0.1}$ & $7.75^{+0.04}_{-0.04}$ & $1.1^{+0.1}_{-0.09}$ & $0.6^{+0.1}_{-0.1}$ & $0.43^{+0.07}_{-0.09}$ & 0.6 \\
8083-12703 & $2.43\pm0.08$ & $8.45\pm0.02$ & $-0.63^{+0.08}_{-0.09}$ & $9.50^{+0.06}_{-0.06}$ & $1.4^{+0.3}_{-0.2}$ & $0.5^{+0.3}_{-0.1}$ & $0.19^{+0.02}_{-0.02}$ & 63.0 \\
8085-6104 & $1.1\pm0.2$ & $7.94\pm0.08$ & $-0.77^{+0.07}_{-0.08}$ & $9.27^{+0.04}_{-0.03}$ & $1.9^{+0.3}_{-0.3}$ & $0.7^{+0.1}_{-0.2}$ & $0.17^{+0.03}_{-0.02}$ & 23.3 \\
8086-3704 & $1.67\pm0.07$ & $8.73\pm0.02$ & $-0.70^{+0.06}_{-0.07}$ & $9.71^{+0.03}_{-0.03}$ & $3.0^{+0.4}_{-0.5}$ & $0.73^{+0.07}_{-0.1}$ & $0.06^{+0.01}_{-0.01}$ & 14.3 \\
8655-1902 & $2.0\pm0.1$ & $8.28\pm0.03$ & $-6^{+3}_{-10}$ & $9.35^{+0.06}_{-0.07}$ & $1.0^{+0.3}_{-0.1}$ & $0.3^{+0.3}_{-0.1}$ & $0.1^{+0.1}_{-0.06}$ & 7.8 \\
8655-3701 & $3.40\pm0.04$ & $9.543\pm0.005$ & $-0.2^{+0.1}_{-0.1}$ & $10.28^{+0.05}_{-0.04}$ & $1.4^{+0.2}_{-0.1}$ & $0.4^{+0.2}_{-0.1}$ & $0.16^{+0.02}_{-0.03}$ & 140.1 \\
8941-3701 & $1.02\pm0.04$ & $8.20\pm0.02$ & $-1.4^{+0.1}_{-0.1}$ & $9.67^{+0.02}_{-0.02}$ & $1.55^{+0.09}_{-0.09}$ & $0.90^{+0.05}_{-0.1}$ & $0.09^{+0.02}_{-0.02}$ & 4.8 \\
8982-6104 & $0.61\pm0.06$ & $8.18\pm0.04$ & $-1.6^{+0.1}_{-0.1}$ & $9.3^{+0.04}_{-0.05}$ & $1.3^{+0.2}_{-0.2}$ & $0.3^{+0.1}_{-0.08}$ & $0.10^{+0.03}_{-0.02}$ & 10.6 \\
9088-9102 & $24.59\pm0.08$ & $10.481\pm0.001$ & $-0.1^{+0.2}_{-0.2}$ & $11.33^{+0.04}_{-0.05}$ & $1.44^{+0.09}_{-0.3}$ & $0.15^{+0.03}_{-0.03}$ & $0.03^{+0.01}_{-0.01}$ & 628.4 \\
9194-3702 & $22.1\pm0.3$ & $10.392\pm0.005$ & $0.77^{+0.08}_{-0.07}$ & $11.0^{+0.04}_{-0.04}$ & $0.81^{+0.08}_{-0.07}$ & $0.3^{+0.08}_{-0.05}$ & $0.31^{+0.07}_{-0.05}$ & 301.1 \\
9494-3701 & $0.96\pm0.05$ & $7.65\pm0.02$ & $-3.1^{+0.6}_{-2.0}$ & $9.73^{+0.02}_{-0.02}$ & $0.67^{+0.03}_{-0.03}$ & $0.26^{+0.03}_{-0.03}$ & $0.31^{+0.03}_{-0.02}$ & 12.0 \\
\end{tabular}
\caption{CO and spectral fitted properties of PSB spaxels in our sample. (1) MaNGA Plate-IFU identifier. (2) CO flux in PSB spaxels. (3) Molecular gas mass in PSB spaxels. (4) Spectral fitted star-formation rate in PSB spaxels. (5) Stellar mass in PSB spaxels. (6) Lookback time to the peak of the starburst. (7) Fraction of stellar mass formed in starburst. (8) Fraction of stellar mass formed in the last Gyr. (9) Inclination corrected surface area of all PSB spaxels.}
\label{tab:region}
\end{table*}

\section{Results} \label{sec:results}

\subsection{Molecular gas morphology}

We consider the morphologies of the integrated CO intensity maps shown in Figure~\ref{fig:overview_co}.
Qualitatively, 6/12 of our PSBs have molecular gas morphologies consisting of one central clump with little extended gas at our resolution.
At least four of the PSBs with extended gas appear to be tidally disturbed in the optical SDSS imaging: 8083-12703, 8655-3701, 9194-3702, and 9088-9102.
The first three are visually classified as post-mergers by \citet{thorp_spatially_2019}, and the latter is an on-going merger, as seen in the SDSS image.
9088-9102 has many high significance CO line detections separated from the galaxy itself, likely due to a plethora of gas in a dense environment.
9494-3701 has extended gas perpendicular to the disc of the galaxy, potentially indicative of gas expulsion or an outflow.

\begin{figure}
    \centering
    \includegraphics[width=0.4\textwidth]{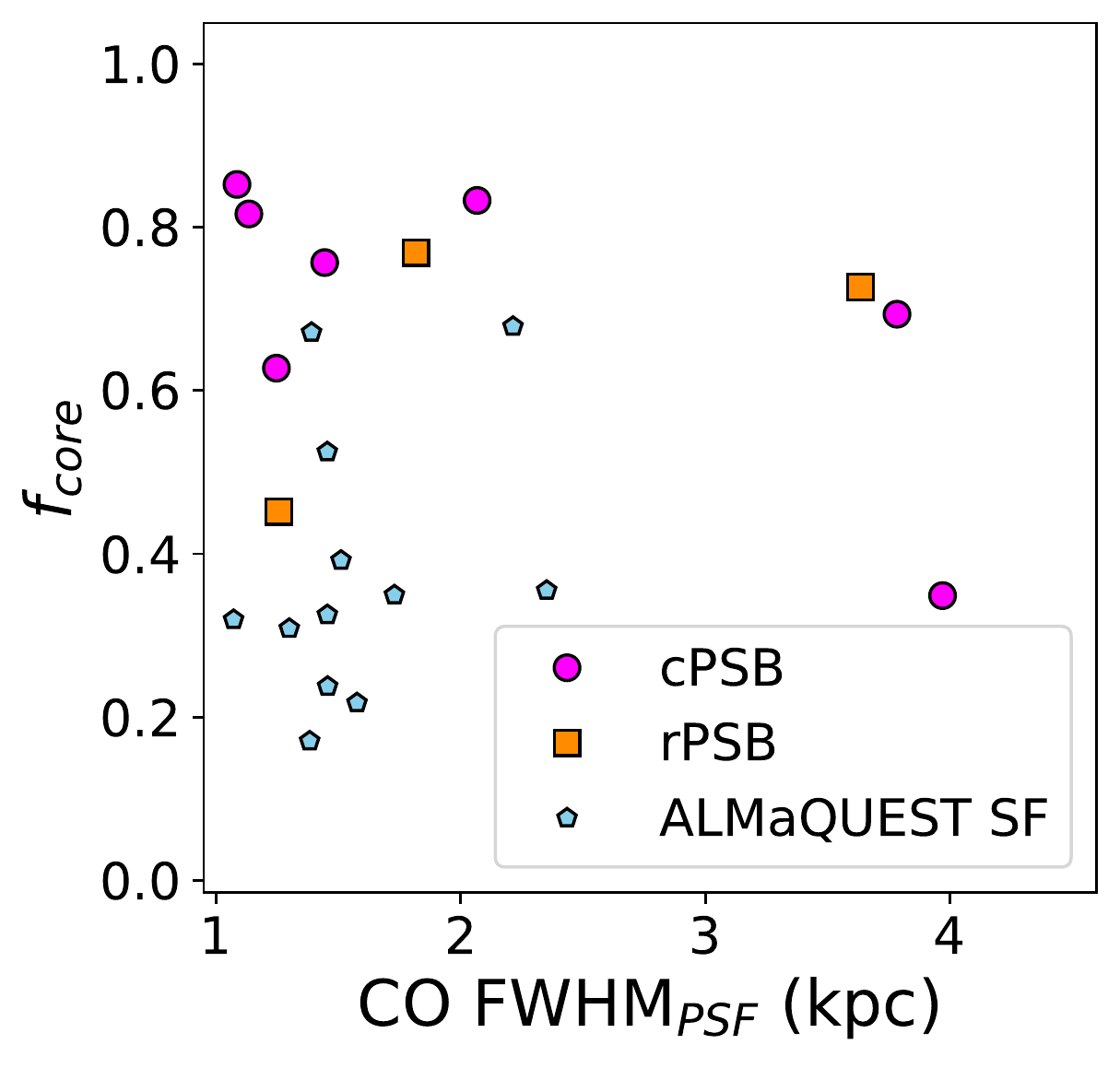}
    \caption{The fraction of total CO line flux within the 3$\sigma$ beam size, $f_{core}$, and the physical resolution of the 2.5\arcsec\, CO beam for our cPSBs in purple, rPSBs in orange, and ALMaQUEST star-forming galaxies in blue. All galaxies plotted were imaged with the same 2.5\arcsec\, circular beam. Our PSBs tend to have higher core fractions and thus more centrally concentrated CO emission even at similar physical resolutions.}
    \label{fig:fcore_psf}
\end{figure}

We fit each CO moment 0 map with a 2D Gaussian.
We do not fit two galaxies: 8081-3702 lacks a robust CO detection, and 8085-6104 has two lobes of gas that cannot be adequately represented with a single Gaussian.
Of the remaining 11 galaxies with reasonable Gaussian fits, 4 have major axis sizes consistent with the ALMA beam size within 20\%, and 9/11 have major axis sizes consistent within 50\% of the beam size.
Following \citet{smercina_after_2022}, we compute the fraction of CO flux within a central core, $f_{core}$, defined as 3$\sigma$ of the ALMA beam size centered on the Gaussian fit center.
We find that 8/11 galaxies have core fractions greater than 50\%, and the sample has a mean core fraction of 66\%.
These high core fractions show that the molecular gas reservoirs in our PSBs are highly concentrated with unresolved centers, consistent with the results of \citet{smercina_after_2022} who also find high core fractions in PSBs with even higher resolution CO observations.
We plot the core fractions versus the physical resolutions of our galaxies in Figure~\ref{fig:fcore_psf}.

We also compute the core fractions of star-forming ALMaQUEST galaxies, finding 3/12 galaxies have core fractions greater than 50\%, and a mean core fraction of 38\%.
We compare the core fraction distributions of the ALMaQUEST star-forming galaxies and our PSBs with a Kolmogorov-Smirnov (KS) test and find a p-value $p = 0.006$, thus the distributions are significantly different.

Overall, at our kpc-scale resolution, our PSBs have more centrally concentrated molecular gas reservoirs than the ALMaQUEST star-forming galaxies.
However, we note that the sample of star-forming galaxies tend to have higher stellar masses and larger optical sizes than our PSBs (see Figure~\ref{fig:context}), leading to larger physical sizes of central gas reservoirs.
Regardless, the central molecular gas reservoirs of our PSBs are compact with sizes $\lesssim$ 1 kpc.
To fully probe the concentration of the central reservoirs of molecular gas, sub-kpc scale resolution observations are needed.

\subsection{Gas and stellar kinematics} \label{sec:kinematics}

\begin{figure*}
    \centering
    \includegraphics[width=\textwidth]{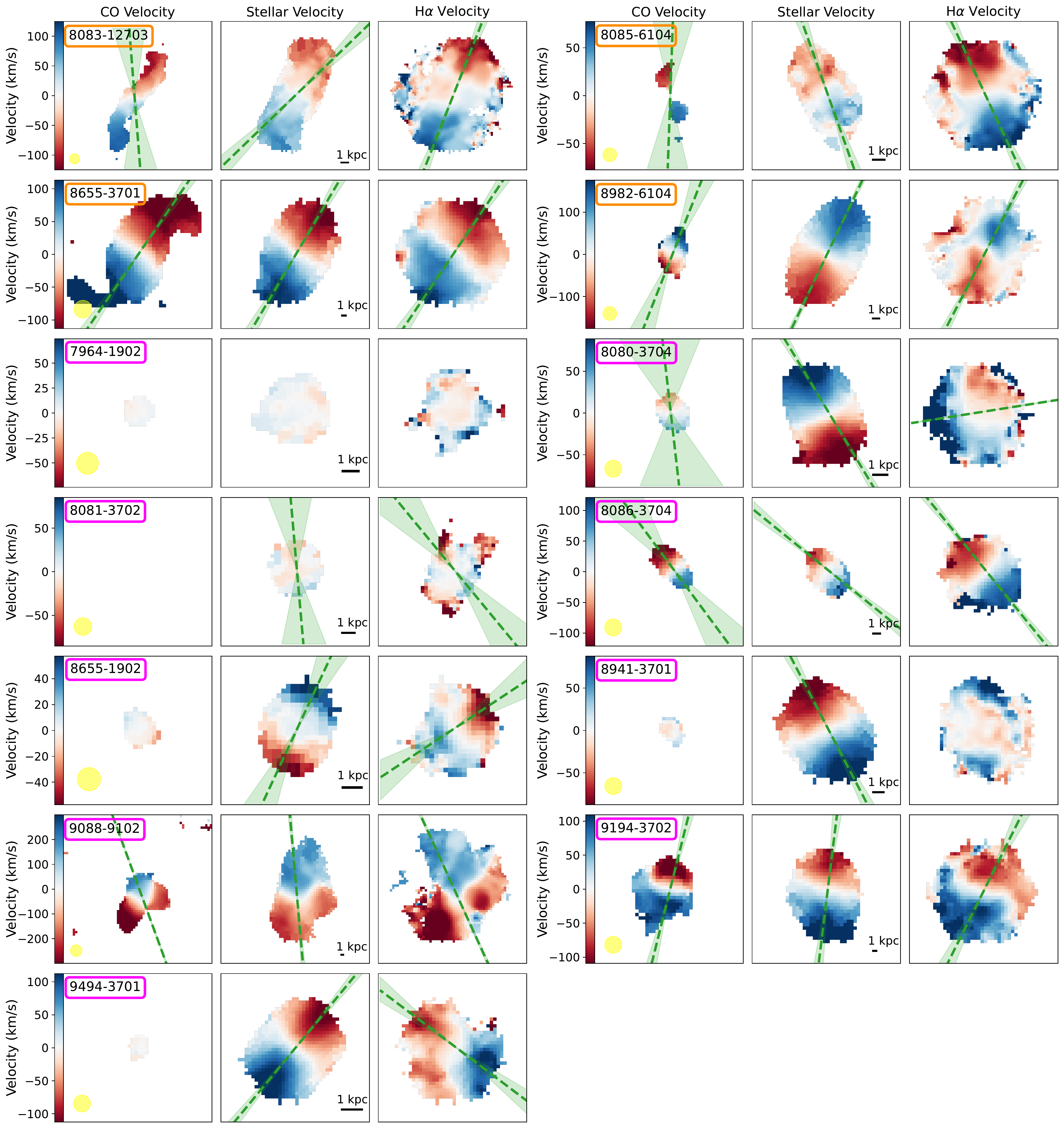}
    \caption{CO, stellar, and H$\alpha$ velocity maps for each galaxy.
    The MaNGA Plate-IFU identifier is in the top left of the CO velocity map, boxed in magenta for central PSBs and orange for ring PSBs.
    The ALMA and MaNGA beam are represented by the yellow ellipse in the bottom left.
    The CO velocity map is shown where we have CO S/N $>$ 3.
    The stellar velocity and H$\alpha$ maps are only shown for spaxels with a S/N of 5 or an H$\alpha$ S/N of 3.
    We fit each velocity map and plot the rotational position angle in green for maps with reliable PA fits, with shaded 3$\sigma$ uncertainties.}
    \label{fig:vmaps}
\end{figure*}

In Figure~\ref{fig:vmaps}, we plot the CO, stellar, and H$\alpha$ velocity maps for each galaxy.
The CO velocity map is described in Section~\ref{sec:obs}, and is shown only for spaxels with CO S/N $>$ 3. 
We also require a g-band S/N $>$ 5 per spaxel for the stellar velocity map, and H$\alpha$ S/N $>$ 3 per spaxel for the H$\alpha$ velocity map.

\begin{table*}[htb!]
    \centering
    \begin{tabular}{c|c|c|c|c|c}
    Plate-IFU (1) & CO PA ($^\circ$) (2) & Stellar PA ($^\circ$) (3) & H$\alpha$ PA ($^\circ$) (4) & Stellar Radon Class (5) & H$\alpha$ Radon Class (6)\\ \hline
    7964-1902 & - & - & - & A & N \\
    8080-3704 & $10\pm30$ & $31\pm3$ & $99.0\pm0.5$ & C & A \\
    8081-3702 & - & $0\pm20$ & $40\pm10$ & N & N \\
    8083-12703 & $0\pm10$ & $134\pm5$ & $158\pm4$ & A & IB \\
    8085-6104 & $180\pm20$ & $20\pm4$ & $25\pm4$ & IB+OB & C \\
    8086-3704 & $40\pm10$ & $51\pm4$ & $39\pm3$ & IB & C \\
    8655-1902 & - & $160\pm10$ & $120\pm10$ & A & C \\
    8655-3701 & $146\pm4$ & $150\pm4$ & $146\pm4$ & A & IB+OB \\
    8941-3701 & - & $28\pm4$ & - & C & A \\
    8982-6104 & $158\pm8$ & $154\pm2$ & $152\pm3$ & IB & IB \\
    9088-9102 & $20.0\pm0.5$ & $5\pm2$ & $24\pm1$ & N & A \\
    9194-3702 & $166\pm4$ & $173\pm4$ & $154\pm5$ & A & A \\
    9494-3701 & - & $140\pm3$ & $53\pm6$ & C & C \\
    \end{tabular}
    \caption{(1) MaNGA Plate-IFU. (2) CO global position angle. Uncertainties are 3$\sigma$ errors for all position angles. PAs are defined counterclockwise from the y-axis. (3) Stellar global position angle. (4) H$\alpha$ global position angle. (5) Stellar Radon profile classification. N = No classification, A = Asymmetric, C = Constant, IB = Inner Bend, OB = Outer Bend, IB+OB = Inner Bend + Outer Bend. (6) H$\alpha$ Radon profile classification, with the same abbreviations.}
    \label{tab:pa}
\end{table*}

We measure the position angle (PA) of each velocity map with the python code \texttt{PaFit}\footnote{\url{https://www-astro.physics.ox.ac.uk/~cappellari/software/\#pafit}} from Appendix C of \citet{krajnovic_kinemetry_2006}, using the optical center of the galaxy.
We only show PAs with 3$\sigma$ uncertainties less than 45$^\circ$.
Finally, we subtract the measured systemic velocities from each velocity field.
We present the PAs for the CO, stellar, and H$\alpha$ velocity fields in Table~\ref{tab:pa}.
We measure the same PAs for the sample of 12 ALMaQUEST star-forming galaxies.
By eye, a few of the PA fits do not appear to follow the velocity field.
This may be the result of imposing a global PA on a velocity field which has a changing PA with radius, as can be seen in the CO emission of 8083-12703. 

We find that 5/12 CO-detected PSBs (2 of which are rPSBs) have CO, stellar, and H$\alpha$ velocity fields with consistent (within 30$^\circ$) PAs.
Other PSBs either lack significant CO or H$\alpha$ rotation (5/13, all cPSBs), or have kinematic misalignments (3/12).
4/12 of our PSBs lack CO rotation entirely, though this may be in part driven by the compactness of the CO detections at our resolution.
We also find that the H$\alpha$ PAs do not always match the CO PAs, with only 5/8 galaxies with reliable PA measurements for both velocity fields agreeing.
In contrast, all 12 of the ALMaQUEST star-forming galaxies have consistent CO, stellar, and H$\alpha$ PAs.

Visual inspection of many of the velocity fields in Figure~\ref{fig:vmaps} shows that a simple global PA does not capture velocity fields with radially dependent PAs or non-asymmetric motions.
To better characterize the velocity fields, we use the Radon transform as implemented in \citet{stark_radon_2018}. 
In brief, the Radon transform is a non-parametric method of determining the PA as a function of radius.
\citet{stark_radon_2018} use an automated method to classify Radon profiles of velocity fields into 5 classes: constant, asymmetric, inner bend, outer bend, and both inner and outer bends.
The various symmetric but non-constant Radon profile classifications (i.e. inner bend, outer bend, and both inner and outer bends) appear to be driven at least in part by a combination of stellar bars, oval distortions, and disk warps, though each of these physical processes may have different Radon profile signatures.
Asymmetric Radon profiles may be driven by tidal interactions due to mergers or gas infall for the H$\alpha$ Radon profiles.
We visually classify the stellar and H$\alpha$ Radon profiles of our PSBs and the ALMaQUEST star-forming galaxies, shown in Table~\ref{tab:pa}.
The majority of galaxies in the PSB and the ALMaQUEST star-forming samples have Radon classifications, though 1-2 galaxies in each sample cannot be classified due to a lack of rotation or spaxels with sufficient S/N.
We do not compute the Radon profiles of our CO velocity fields as this method requires relatively complete velocity information over a chosen radius, such as $R_e$, which few of our CO velocity fields satisfy.

\begin{figure*}
    \centering
    \includegraphics[width=0.8\textwidth]{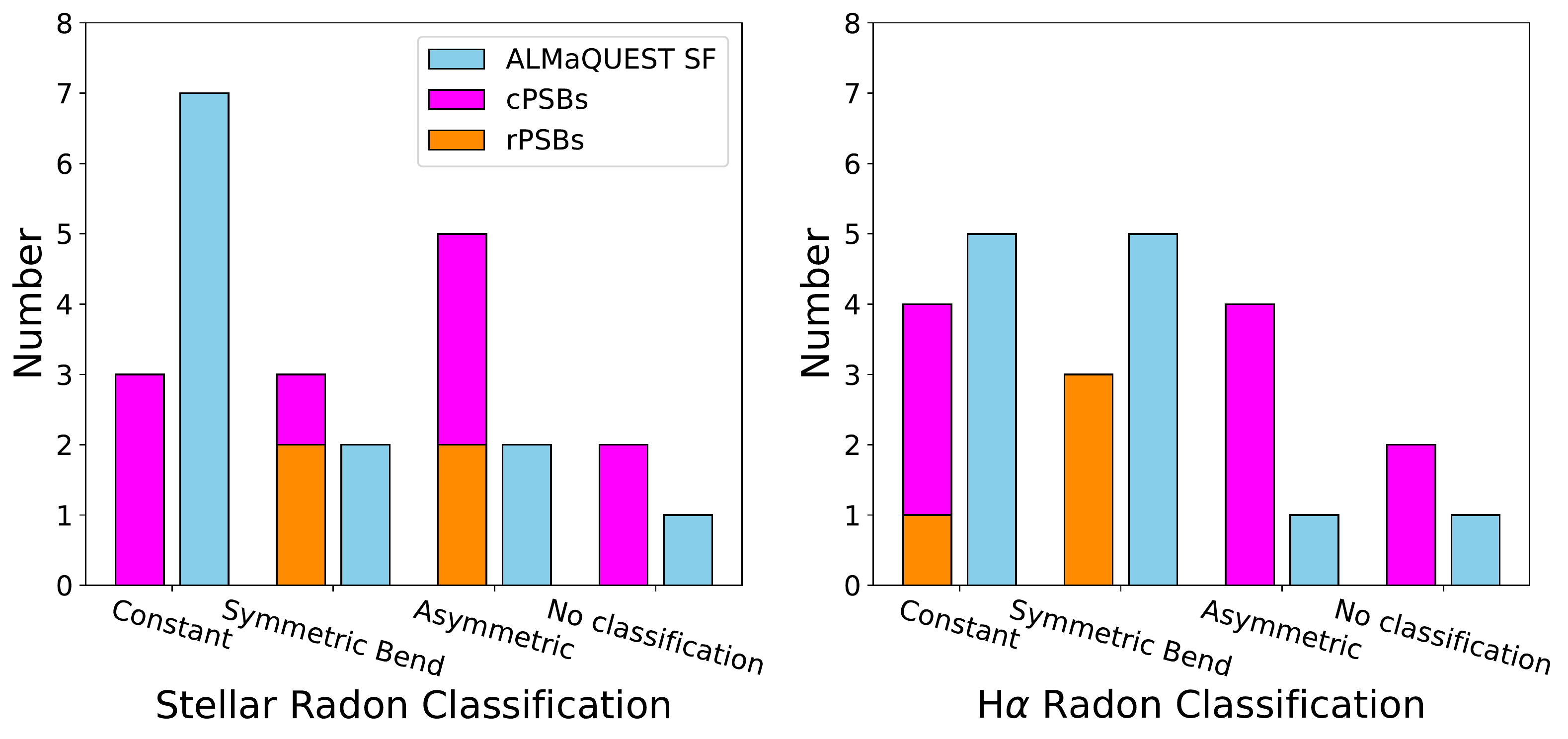}
    \caption{Histograms showing the Radon classifications for stellar velocity fields on the left and H$\alpha$ velocity fields on the right. cPSBs are plotted in purple, rPSBs in orange, and ALMaQUEST star-forming galaxies in blue. Due to our small sample, we combine the inner bend, outer bend, and inner bend + outer bend classifications from \citet{stark_radon_2018} into the symmetric bend category. In both panels, PSBs show an excess of asymmetric velocity fields.}
    \label{fig:radon}
\end{figure*}

Our PSBs are more likely to have asymmetric stellar and H$\alpha$ Radon profiles than the ALMaQUEST star-forming galaxies.
Figure~\ref{fig:radon} shows histograms of the Radon profile classifications for the stellar and H$\alpha$ velocity fields. 
We combine the inner bend, outer bend, and inner bend + outer bend classifications into the symmetric bend category as our sample is too small to meaningfully split into these categories, though we emphasize different physical processes likely drive the different types of bends.
For the stellar (H$\alpha$) Radon profiles, 5/11 (4/11) PSBs have asymmetric profiles, whereas 2/11 (1/11) are asymmetric for ALMaQUEST star-forming galaxies.
While these results are suggestive, with our small sample sizes these differences are not statistically significant.
From \citet{stark_radon_2018}, the percentages of asymmetric stellar and H$\alpha$ profiles in the entire MaNGA DR15 sample are 15.0$^{+1.7}_{-1.5}$\% and 24.4$^{+2.4}_{-2.1}$\% respectively.
Our percentage of asymmetric stellar profiles (45$^{+15}_{-14}$\%) is significantly higher from that of \citet{stark_radon_2018} with a p-value of 0.03, whereas the percentage of asymmetric H$\alpha$ profiles in our sample (36$^{+15}_{-13}$\%) is indistinguishable, with a p-value of 0.19.
Finally, 0/4 rPSBs have asymmetric H$\alpha$ Radon profiles, while 4/8 cPSBs do, but a larger sample is needed to determine if this is a physical difference.

Considering both the stellar and gas kinematics, we can classify our galaxies into three categories: those with asymmetric or non-rotating stellar velocity fields (7/13), those with symmetric stellar rotation but asymmetric or misaligned gas velocity fields (3/13), and those with consistent stellar, CO, and H$\alpha$ rotation (3/13).
In the second category, all three galaxies have constant stellar Radon profiles but either have asymmetric H$\alpha$ Radon profiles as in 8080-3704 and 8941-3701, or are kinematically misaligned as in 9494-3701.
Finally, the galaxies in the last category (8085-6104, 8086-3704, and 8982-6104) have symmetric stellar and H$\alpha$ Radon profiles, as well as consistent CO, stellar, and H$\alpha$ PAs.
We will discuss potential physical causes for these different kinematic patterns in Section~\ref{sec:discuss}.

Overall, the kinematics of our PSBs tend to be more complex than those of the ALMaQUEST star-forming galaxies, with misalignments and asymmetries being more common.
Nearly half of our PSBs (5/12) have CO, H$\alpha$, and stellar velocity fields rotating with a consistent position angle, whereas all ALMaQUEST star-forming galaxies (12/12) have consistent PAs.
Finally, we use the Radon transform to study the rotating stellar and H$\alpha$ velocity fields and find that asymmetric Radon profiles appear to be more common in PSBs than the ALMaQUEST star-forming galaxies for both stellar and H$\alpha$ velocity fields.

\subsection{Star-formation and molecular gas properties in context}

\subsubsection{Global properties}

\begin{figure*}
    \centering
    \includegraphics[width=\textwidth]{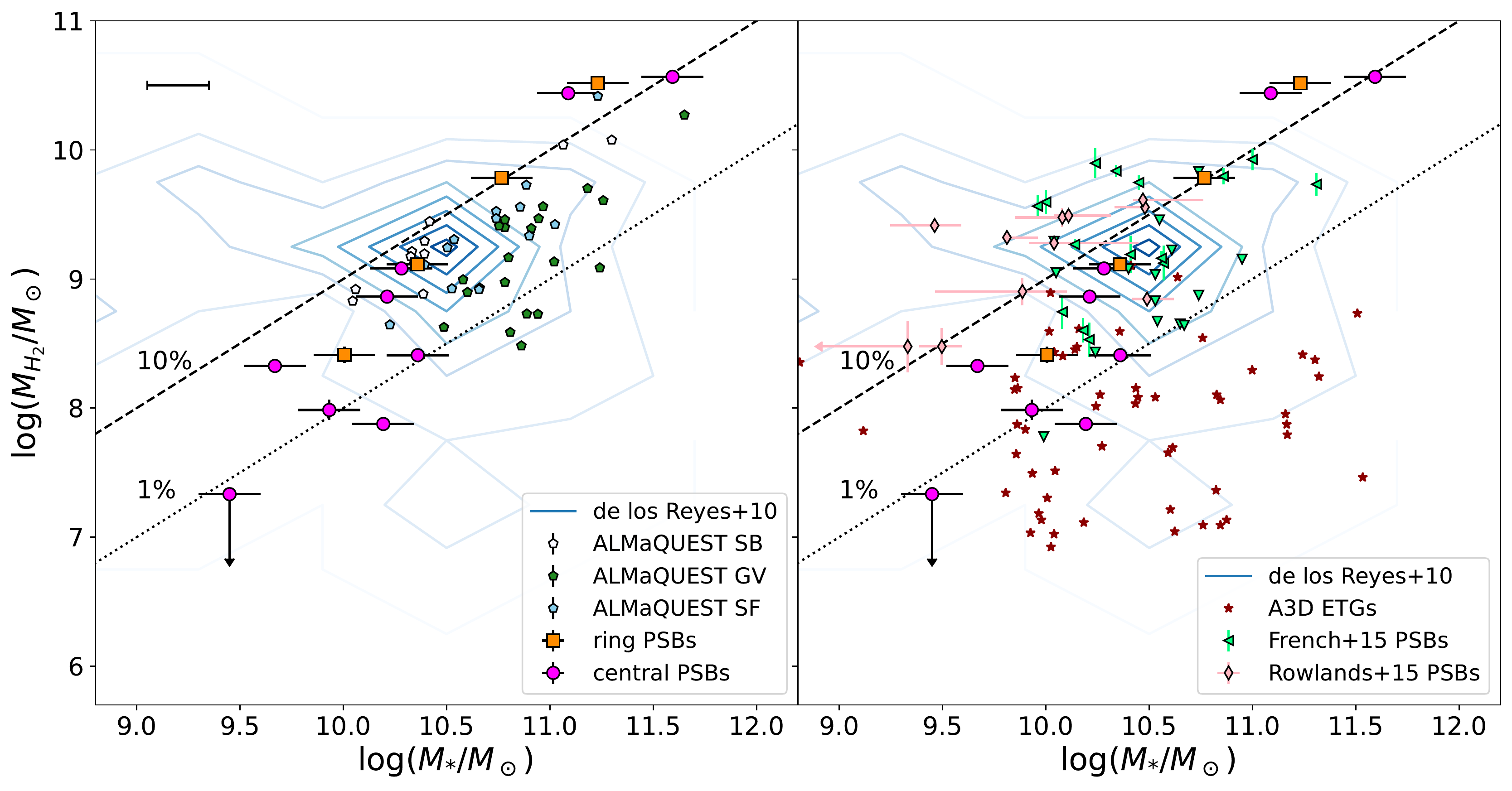}
    \caption{Total molecular gas and stellar masses for our cPSBs and rPSBs in purple and orange respectively, with comparison samples plotted.
    The dotted and dashed lines show constant molecular gas fractions of 1 and 10\% respectively.
    The blue contours show star-forming galaxies from \citet{de_los_reyes_revisiting_2019}.
    In the left panel, the ALMaQUEST star-forming, green valley, and starburst samples are plotted in teal, green and white pentagons respectively, with molecular gas masses and star-formation rates summed within 1.5 $R_e$ of each galaxy. The errorbar in the upper left shows typical uncertainties for the ALMaQUEST galaxies.
    In the right panel, two unresolved studies of PSBs are shown, \citet{french_discovery_2015}, a sample of E+A galaxies in light green triangles, and \citet{rowlands_evolution_2015} in pink diamonds, a sample of PSBs spanning a range of post-burst ages.
    Downward green triangles are \citet{french_discovery_2015} E+As with gas mass upper limits.
    Early type galaxies from the ATLAS$^{\text{3D}}$ survey \citep{young_atlas3d_2011} are plotted as dark red stars.
    We use the same $\alpha_{CO} = 4.35$M$_\odot$ (K km s$^{-1}$ pc$^2$)$^{-1}$ for all samples plotted.
    Our PSBs have a wide range of molecular gas fractions; while some of our PSBs are very gas-rich, others have little molecular gas.
    }
    \label{fig:mgas_global}
\end{figure*}

Previous single-dish CO observations of PSBs have reported large molecular gas fractions despite the low star-formation rates of these galaxies \citep{french_discovery_2015, rowlands_evolution_2015, alatalo_shocked_2016-1}.
In Figure~\ref{fig:mgas_global}, we plot the total molecular gas masses and the total stellar masses from \citet{pace_resolved_2019} for our sample and a number of comparison samples.
In both panels we plot the \citet{de_los_reyes_revisiting_2019} sample of star-forming galaxies as contours.
On the left, we plot the three ALMaQUEST samples: star-forming galaxies (SF), green valley galaxies (GV), and starburst galaxies  \citep[SB,][]{lin_almaquest_2019, lin_almaquest_2020, ellison_starburst_2020}.
On the right, we compare our PSBs to unresolved CO observations of PSBs, including the samples of \citet{french_discovery_2015} and \citet{rowlands_evolution_2015}, as well as early-type galaxies from the ATLAS$^{\text{3D}}$ survey \citep{young_atlas3d_2011, cappellari_atlas3d_2011}.
For all the plotted samples, we use a constant $\alpha_{CO}$ of 4.35 M$_\odot$ (K km s$^{-1}$ pc$^2$)$^{-1}$ for consistency.
This value of $\alpha_{CO}$ is likely an overestimate for the SB galaxies, so gas masses for this sample can be considered upper limits \citep{bolatto_co--h2_2013}.

Our PSBs span a range of global molecular gas fractions, from $<$1\% to above 20\%.
While some of our PSBs have gas fractions similar to or greater than the majority of the \citet{de_los_reyes_revisiting_2019} star-forming sample, 4 of our cPSBs have low gas fractions $\lesssim$ 1\%.
Other plotted molecular gas studies of PSBs similarly show a wide range of molecular gas fractions, though the greater sensitivity of our observations allows us to measure lower gas fractions down to 1\%.
We also see that our PSBs tend to have more molecular gas than the early-type galaxies, especially for PSBs with $\log M_*/M_\odot > 10.5$.

\subsubsection{Post-starburst regions}

\begin{figure}
    \centering
\includegraphics[width=0.45\textwidth]{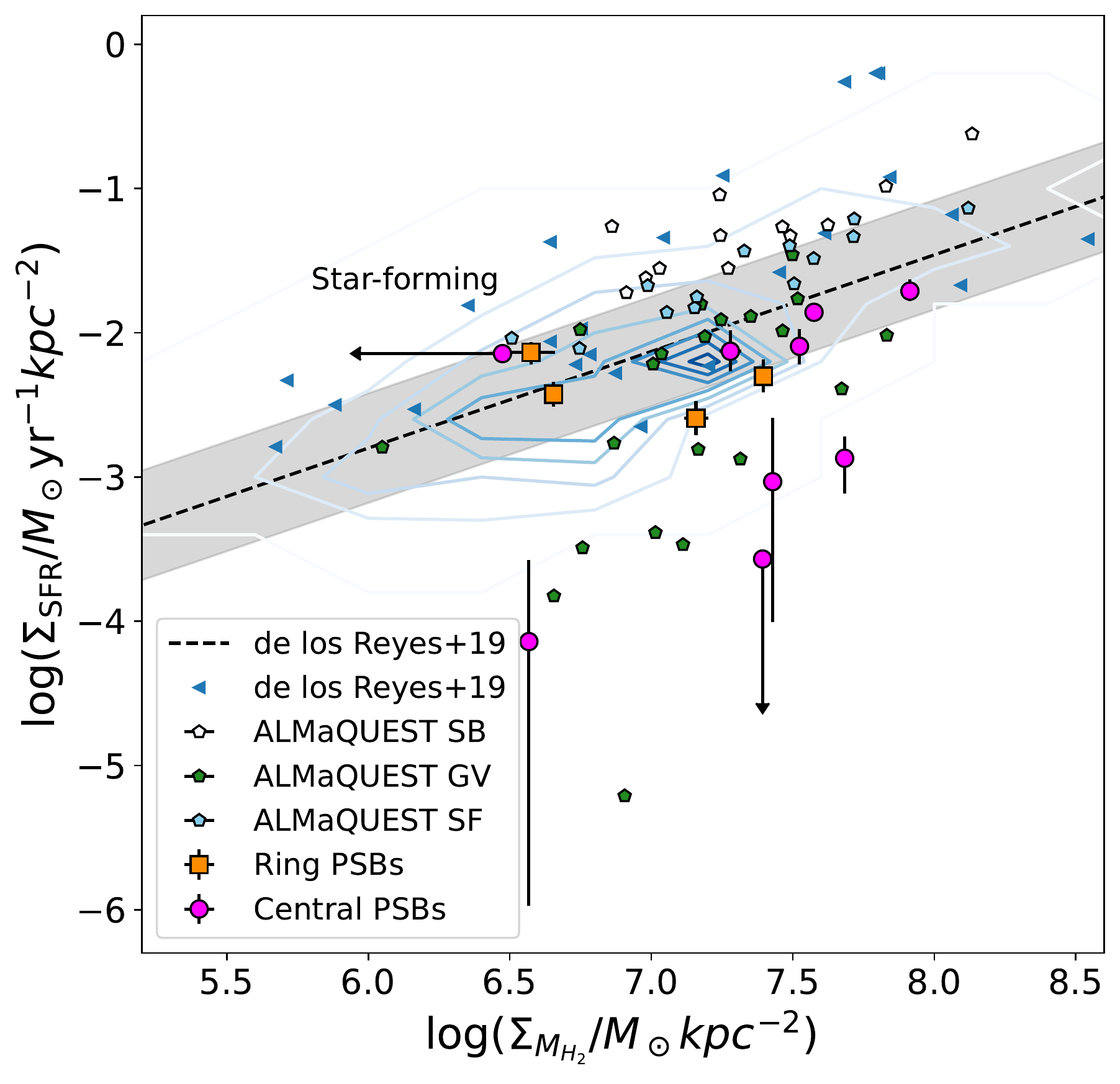}
    \caption{The Kennicutt-Schmidt relation between star-formation surface density and molecular gas surface density for post-starburst regions in our sample in purple for cPSBs and orange for rPSBs.
    We plot the ALMaQUEST star-forming, green valley, and starburst galaxies as blue, green, and white pentagons respectively.
    The dashed line shows the measured relation from \citet{de_los_reyes_revisiting_2019} with the estimated scatter shaded as gray.
    }
    \label{fig:ks}
\end{figure}

The link between star-formation surface density and molecular gas surface density, the Kennicutt-Schmidt relation, is empirically supported in a variety of studies and physically motivated \citep{kennicutt_jr_global_1998, de_los_reyes_revisiting_2019}. 
In Figure~\ref{fig:ks}, we plot this relation for the post-starburst regions of our galaxies compared to a variety of other studies.
We plot the properties of only the post-starburst regions of our galaxies rather than global properties to investigate the star-formation properties of the post-starburst regions specifically.
Given that the kpc-scale post-starburst regions have physical sizes much larger than an average giant molecular cloud for each galaxy, local small-scale variations in the molecular gas distribution will not impact our results.
To compute the post-starburst region surface densities, we use the spectral fitted SFR which was measured over the entire post-starburst region and sum the molecular gas masses over the post-starburst region.
We divide these by the area of the post-starburst regions, which we measure by multiplying the number of post-starburst spaxels by their inclination-corrected physical surface areas, with inclination values from the NSA.
As shown in Figure~\ref{fig:fcore_psf}, we are unable to resolve the central molecular gas, thus our measured molecular gas surface densities are effectively lower limits (at least in our cPSBs) as the gas could be significantly more compact than our beam size, as seen in \citet{smercina_after_2022, luo_ic860_2022}.

In Figure~\ref{fig:ks}, we plot the star-formation rate and molecular gas surface densities for our PSBs, the star-forming sample of \cite{de_los_reyes_revisiting_2019}, and the three ALMaQUEST samples.
The ALMaQUEST star-formation rates are computed from the extinction-corrected H$\alpha$ emission according to \citet{kennicutt_jr_global_1998}, but only for spaxels classified as star-forming according to the [SII]/H$\alpha$ vs [OIII]/H$\beta$ diagram \citep{kewley_host_2006}.
\citet{de_los_reyes_revisiting_2019} compute SFRs for their sample with UV fluxes, correcting for dust extinction with IR fluxes.
For typical star-forming galaxies, these two SFR indicators are tightly correlated \citep{hao_dust-corrected_2011}.
For the GV and SB sample, the H$\alpha$ SFRs may be underestimated in galaxies with large numbers of spaxels with AGN or composite classified spaxels, as any star-formation from these spaxels is neglected.

Compared to the local star-forming galaxies of \citet{de_los_reyes_revisiting_2019}, we see that 5/13 of our PSB regions lie significantly below the fitted relation, thus having lower star-formation efficiencies.
In particular, 8655-1902 and 8941-3701 are two central PSBs with sizable molecular gas reservoirs that have nearly quenched their star-formation entirely.
Other post-starburst regions are consistent with the Kennicutt-Schmidt relation of \citet{de_los_reyes_revisiting_2019}.
Our sample is offset below the star-forming (SF) and starburst (SB) ALMaQUEST samples with lower star-formation surface densities as expected, and occupies a similar region as the green-valley (GV) galaxies, though often with lower gas surface densities.

\begin{figure}
    \centering
    \includegraphics[width=0.45\textwidth]{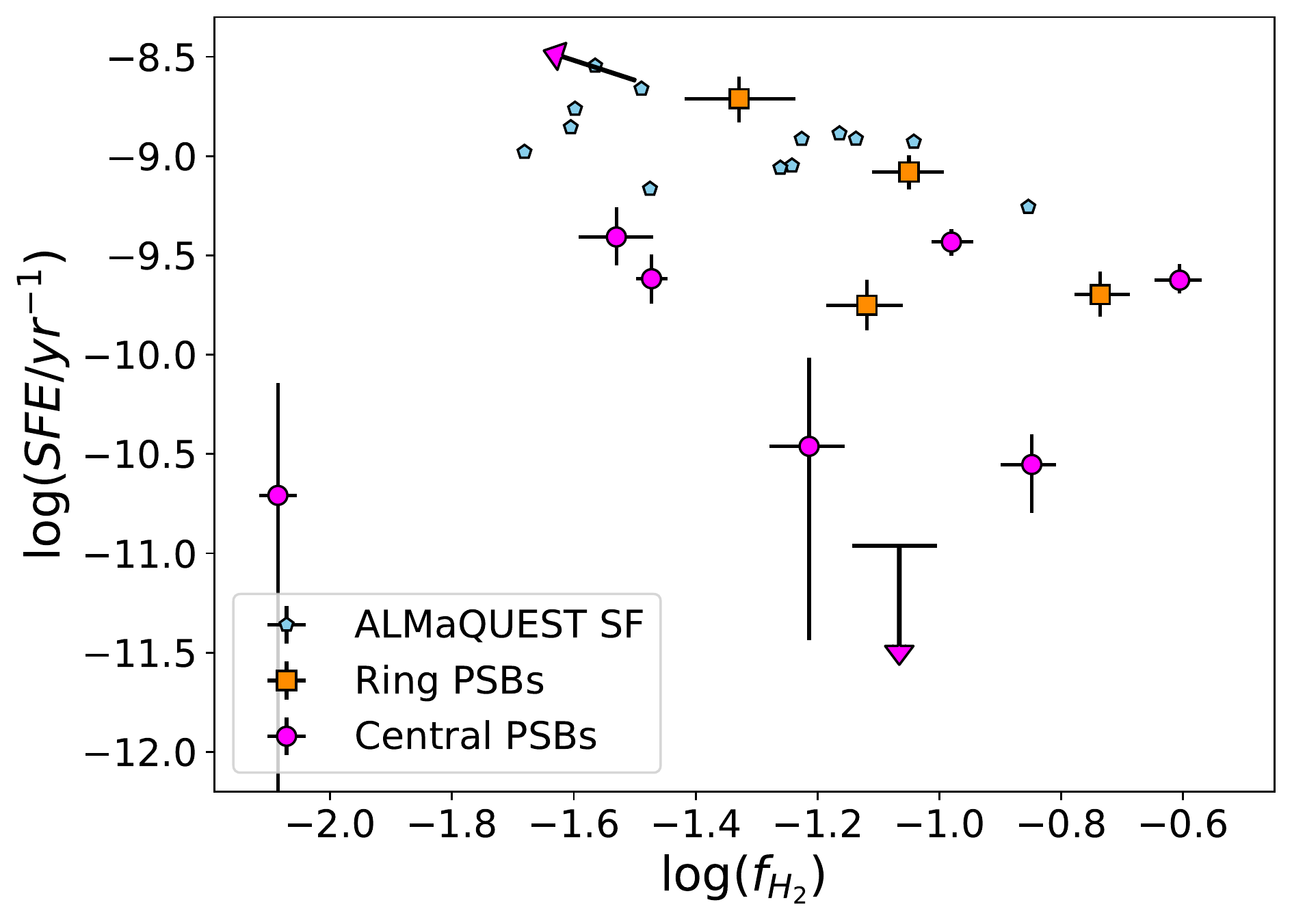}
    \caption{The star-formation efficiency (SFR / $M_{H_2}$) and molecular gas fraction ($M_{H_2} / M_*$) for post-starburst regions (central PSBs in purple, and ring PSBs in orange) and the ALMaQUEST star-forming galaxies. The downward arrow shows an upper limit in star-formation rate, and the diagonally pointing arrows show an upper limit in $M_{H_2}$ with constant SFR and $M_*$.}
    \label{fig:sfe_fgas}
\end{figure}

In Figure~\ref{fig:sfe_fgas}, we plot the star-formation efficiency (SFE = $\Sigma_{\textrm{SFR}}$ / $\Sigma_{M_{H_2}}$) and the molecular gas fraction ($f_{H_2}$ = $M_{H_2}$ / $M_*$) of the PSB regions and the ALMaQUEST star-forming galaxies.
The majority of our PSBs have similar $f_{H_2}$ as the star-forming galaxies but have suppressed SFEs, though the degree of this suppression varies significantly.
One cPSB, 9494-3701, has a particularly low $f_{H_2}$, potentially indicating that a lack of molecular gas supply is contributing to quenching in this galaxy.
We are considering the post-starburst regions of our galaxies rather than global properties, thus \textit{within} the quenching post-starburst region we find that there is typically still a significant amount of gas forming stars at a suppressed rate.
As discussed above, the molecular gas mass surface densities for our post-starburst regions are effectively lower limits, and therefore the plotted SFEs are upper limits.

\subsection{Molecular gas and the star-formation history}

\begin{figure*}
    \centering
    \includegraphics[width=\textwidth]{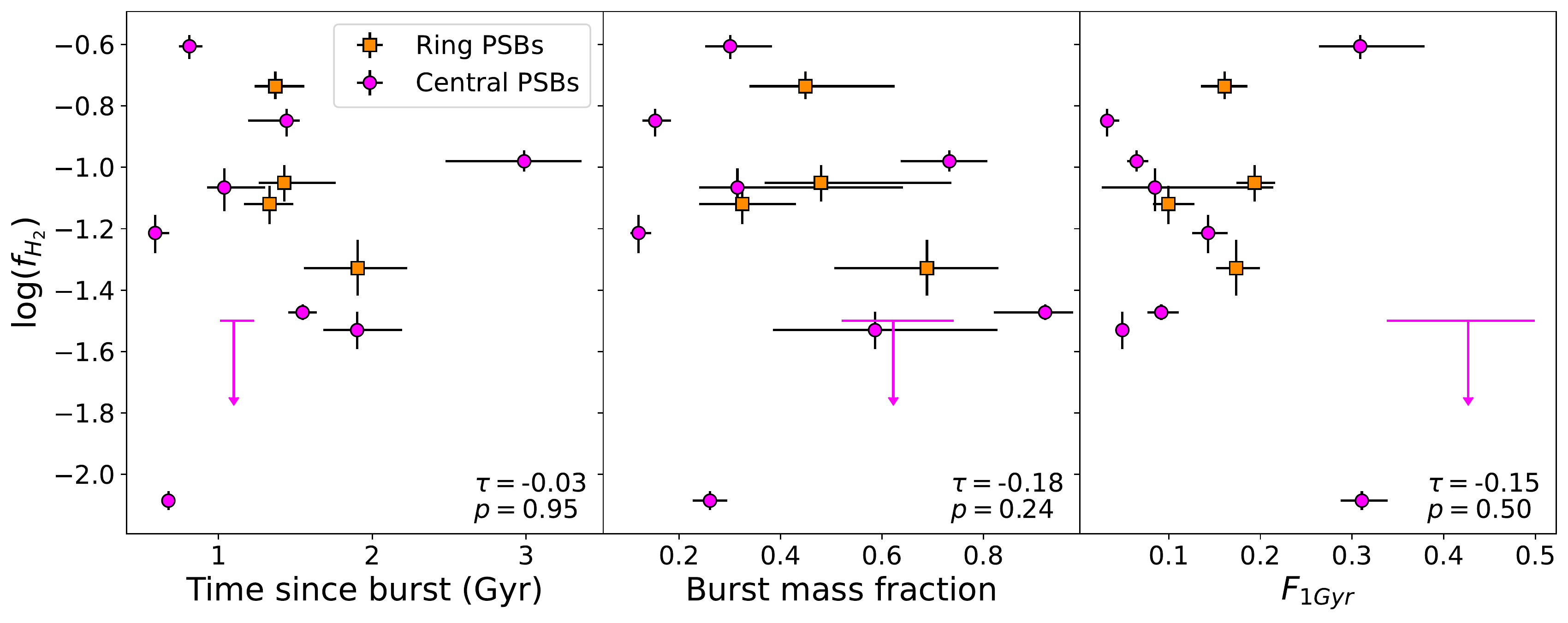}
    \caption{Correlations between the log molecular gas fraction ($M_{H_2}$/$M_*$) of PSB regions and star-formation history parameters.
    cPSBs are plotted in purple and rPSBs in orange.
    Kendall's $\tau$ correlation coefficient (including upper limits) and accompanying p-value are shown in the bottom right of each panel.
    We do not observe a significant correlation between any of the plotted quantities.
    Left: log molecular gas fraction  versus the time since the burst, approximately the time since the peak of the starburst.
    Center: log molecular gas fraction and the fraction of stellar mass formed in the recent starburst.
    Right: log molecular gas fraction and the fraction of stellar mass formed in the last Gyr.}
    \label{fig:fgas_sfh}
\end{figure*}

Observational and theoretical studies have found a negative correlation between the post-burst age of a PSB and the molecular gas fraction \citep{rowlands_evolution_2015, french_clocking_2018, davis_evolution_2019}.
The timescale of the removal of molecular gas reservoirs after the starburst can provide clues about which physical mechanisms are driving this removal, such as AGN-driven outflows. 
We plot the molecular gas fraction as a function of the time since starburst, burst mass fraction, and fraction of stellar mass formed in the last Gyr ($F_{1Gyr}$) for our PSB regions in Figure~\ref{fig:fgas_sfh}.
The time since starburst is the lookback time when the peak of the burst occurs, consistent with the definition of $t_{\text{burst}}$ in Equation~\ref{eq:psb_wild2020} and from \citet{wild_star_2020}.

With our small sample, we do not observe any evolution in molecular gas fraction after the starburst, burst mass fraction, or the fraction of stellar mass formed in the last Gyr.
To determine whether any of these correlations are statistically significant, we use Kendall's $\tau$ rank correlation coefficient, a non-parametric correlation test which typically yields similar results as Spearman's $\rho$.
As discussed in \citet{feigelson_modern_2012}, this coefficient can be generalized to include upper limits, as is necessary with our CO non-detection.
We find no statistically significant correlations between molecular gas fraction and the plotted SFH parameters in Figure~\ref{fig:fgas_sfh}.
However, our sample is small and heterogeneous with both cPSBs and rPSBs, so any correlations may be washed out by the intrinsic scatter of the relation, such as the range of gas fractions before the post-starburst phase.
We might expect a correlation between the gas fraction and the burst mass fraction and $F_{1Gyr}$ if gas removal is driven by an exhaustion of gas reservoirs in the starburst.
However, we observe no such correlation, though this may again be due to the small size of our sample.

To determine whether the overall ISM content of our PSBs is evolving, we attempt a similar analysis with HI gas mass measurements from the HI-MaNGA survey \citep{stark_h_2021} in Appendix~\ref{app:HI}, but are unable to draw conclusions due to a lack of HI measurements with low source confusion probabilities in our sample.

\subsection{Emission line diagnostics} \label{sec:emlines}

\begin{figure*}
    \centering
    \includegraphics[width=0.88\textwidth]{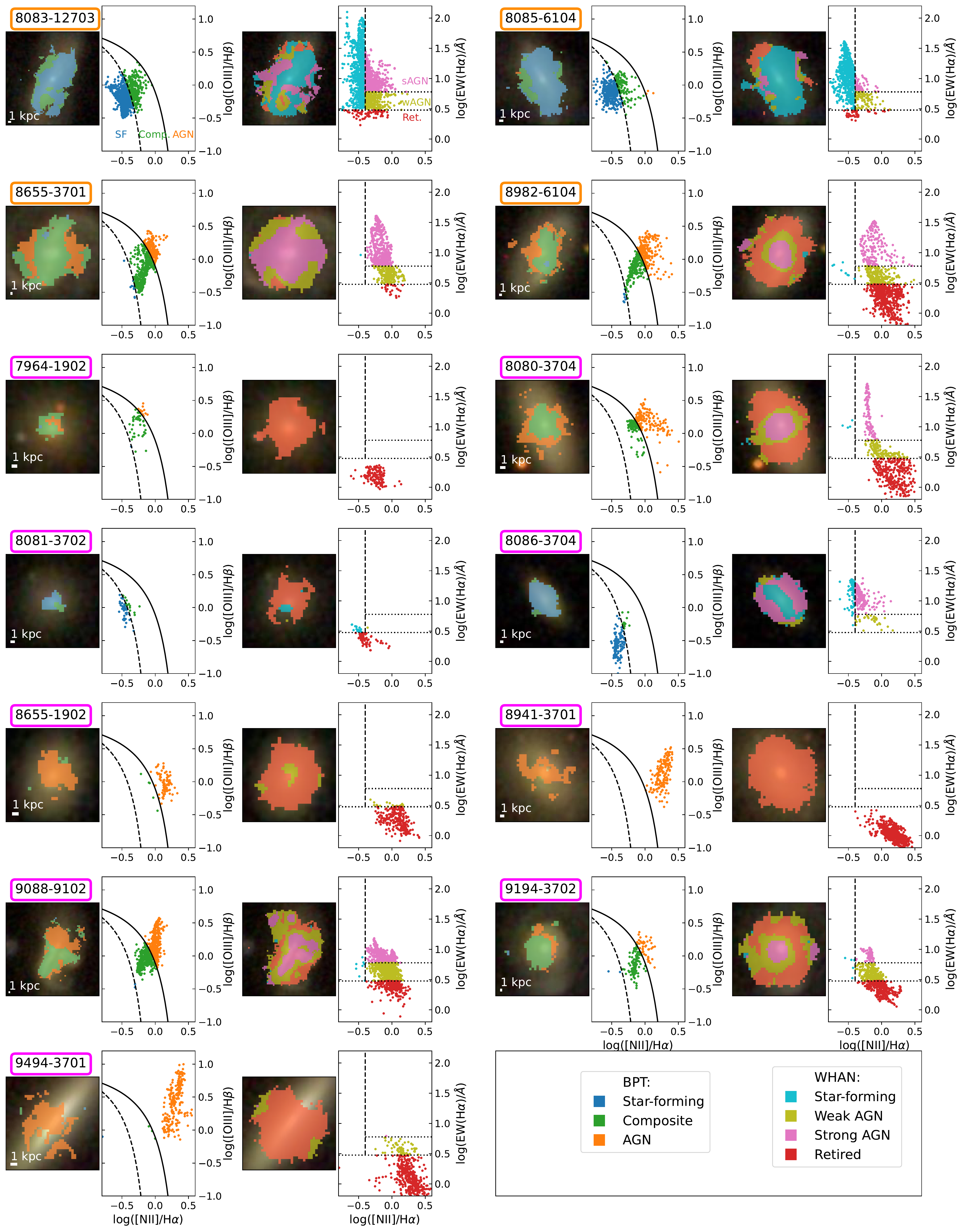}
    \caption{Spatially resolved [NII] BPT diagrams and WHAN diagrams for each galaxy. Left: SDSS 3-color image with composite, AGN, and star-forming classified spaxels in green, orange, and blue respectively, next to the [NII]/H$\alpha$ versus [OIII]/H$\beta$ diagram for each galaxy with the same coloring. The solid black line is from \citet{kewley_theoretical_2001}, and the dashed line is from \citet{kauffmann_host_2003}. Right: SDSS 3-color image with star-forming, weak AGN, strong AGN, and retired classified spaxels in teal, brown, pink, and red respectively, alongside the WHAN diagram, W(H$\alpha$) versus [NII]/H$\alpha$ with classification boundaries from \citet{cid_fernandes_comprehensive_2011}.
    }
    \label{fig:overview_bpt}
\end{figure*}

We classify the ionization source of the gas in our galaxies using the Baldwin, Phillips, and Terlevich diagram \citep[BPT diagram][]{baldwin_classification_1981} and the W(H$\alpha$) versus [NII]/H$\alpha$ (WHAN) diagram \citep{cid_fernandes_comprehensive_2011}.
Figure~\ref{fig:overview_bpt} shows both diagrams for each galaxy, along with the ionization source classifications overlaid on the SDSS 3-color image for both diagrams.
We require each line to have S/N $\geq$ 3.
We classify AGN spaxels with the BPT diagram as above the \citet{kewley_theoretical_2001} line, and composite spaxels as below this line but above the \citet{kauffmann_host_2003} line.
In the WHAN diagram, strong and weak AGN spaxels have [NII]/H$\alpha$ $>$ 0.4 while strong AGN spaxels have W(H$\alpha$) $>$ 6 \AA, weak AGN have W(H$\alpha$) $>$ 3 \AA, and retired spaxels have W(H$\alpha$) $<$ 3 \AA.
These retired spaxels are ionized by hot low-mass evolved stars (HOLMES).
With only optical emission lines, it is difficult to unambiguously separate AGN and shock ionization \citep[e.g.][]{alatalo_shocked_2016, kewley_understanding_2019}.
However, given the central location of an AGN, off-center AGN-classified spaxels in both diagrams are likely to be ionized from shocks \citep{kewley_understanding_2019}.

Three galaxies are dominated by star-forming spaxels in both diagrams: 8083-12703, 8085-6104, and 8086-3704.
All three of these galaxies also have off-center BPT composite spaxels and WHAN strong/weak AGN spaxels, a likely sign of shock ionization.
Five galaxies are dominated by retired spaxels in the WHAN diagram, corresponding to AGN/composite BPT spaxels: 7964-1902, 8081-3702, 8655-1902, 8941-3701, and 9494-3701.
While 8655-1902 does have some central weak AGN spaxels, the diagram shows these spaxels are close to the border with the retired classification, making it difficult to differentiate between these two ionization sources.
Finally, the remaining five galaxies have central regions dominated by BPT AGN or composite spaxels and WHAN strong AGN spaxels: 8655-3701, 8982-6104, 8080-3704, 9088-9102, and 9194-3702.
The spatial extent of the AGN ionization varies; 8655-3701 consists of almost entirely WHAN strong and weak AGN spaxels, whereas the other galaxies have a central core of WHAN strong AGN spaxels surrounded by a ring of weak AGN spaxels and then retired spaxels.
Three of these central AGN galaxies also have isolated islands of off-center WHAN strong AGN spaxels: 8982-6104, 9088-9102, and 9194-3702, which again may be ionized from shocks.

Overall, we find that ionization in 4-5/9 our cPSBs is driven by HOLMES, 3-4/9 by AGN, and star-formation in 1/9, where 8655-1902 is an ambiguous case where HOLMES or a weak AGN could be contributing to central ionization.
The percentage of 33-44\% cPSB AGN hosts is broadly consistent with previous work that has found that an central supermassive black hole is an AGN for approximately 50\% of the post-starburst phase.
Finally, 4/9 of our cPSBs show evidence of non-central shock ionization.

As for our rPSBs, 2/4 have central star-formation with non-central shocks, and the other 2/4 are ionized by an AGN.
Our small sample of rPSBs similarly has a high AGN fraction.
The former two galaxies are rPSBs; this ionization pattern is consistent with the centers of these rPSBs being under typical star-forming conditions with shocks or other energetic processes ionizing and disrupting gas in the outskirts.

\section{Discussion} \label{sec:discuss}

\subsection{A tumultuous history for post-starbursts} \label{sec:disc_tumult}

Overall, centrally concentrated molecular gas reservoirs and disturbed stellar kinematics in 7/13 PSBs provide strong evidence that these galaxies have undergone a recent merger.  
For 8/13 of our PSBs, the majority of the CO emission resides in a central core that is unresolved with our kpc-scale resolution CO observations.
Though few other PSBs have spatially resolved CO observations, other studies also reveal highly centrally concentrated molecular gas distributions \citep{smercina_after_2022}. 
Even with higher resolution CO observations (down to $\sim$200 pc spatial resolution) \citet{smercina_after_2022} are still unable to resolve the central cores containing the majority of the molecular gas in a sample of 6 PSBs.
These centrally concentrated CO morphologies are consistent with predictions from \citet{davis_evolution_2019}.
By studying galaxies that have undergone a recent post-starburst phase in the EAGLE simulation, they find that EAGLE PSBs tend to have centrally concentrated molecular gas reservoirs and disturbed kinematics primarily due to mergers disrupting the gas and driving it inwards, sometimes resulting in a starburst or an increase in AGN accretion, both of which deplete molecular gas reservoirs quickly.

Asymmetric or disordered stellar velocity fields further support a merger history for 7/13 of our cPSBs. 
While kinematic features such as stellar bars, oval distortions, and disk warps (which may also be of merger origin) could lead to symmetric alterations to the stellar velocity fields, tidal interactions from mergers are the most likely cause of stellar asymmetries.
These disturbances are not limited to the stellar velocity field, as the molecular gas reservoirs in our 4 of our PSBs lack rotation, indicating that the molecular gas reservoirs themselves are disturbed. 
Of the 12 ALMaQUEST star-forming galaxies, only 2 have stellar velocity fields with asymmetric Radon profiles, and all 12 have consistent CO, stellar, and H$\alpha$ position angles, in comparison to only 5/12 PSBs. 
Thus, the kinematics of our PSBs are less orderly and more disturbed than in the ALMaQUEST star-forming sample, indicating that physical processes disturbing kinematics in our galaxies are more prevalent.
Four of our PSBs are visually classified as either post-mergers (8083-12703, 8655-3701, and 9194-3702) by \citet{thorp_spatially_2019}, or on-going mergers (9088-9102).
The stellar Radon profiles of all four of these galaxies are either asymmetric or could not be classified due to a lack of rotation.

A substantial body of work supports gas-rich mergers leading to a starburst and subsequent quenching episode resulting in a PSB phase.
In multiple simulations, a majority of PSBs have recently undergone mergers, making them an important formation mechanism \citep{davis_evolution_2019, zheng_comparison_2020, lotz_new_2004}.
Observationally, PSBs are morphologically disturbed and show tidal features at higher rates than typical star-forming or early-type galaxies \citep{yang_detailed_2008, sazonova_are_2021}, further supporting a recent merger for at least half of local PSBs.

While mergers may drive the bulk of the observed stellar asymmetries, smaller interactions or outflows may drive star-formation suppression in the 3/12 of our PSBs with symmetrical stellar Radon profiles and misaligned or asymmetric H$\alpha$ Radon profiles.
A minor merger or interaction event could trigger a starburst and the following post-starburst phase \citep{wilkinson_evolution_2018}, which could result in the misalignment of the gas without significantly altering the stellar rotation.
Alternatively, an outflow could drive a disturbance in the gas while leaving the stellar velocity field intact.
Outflows have previously been observed in PSBs \citep[e.g.][]{alatalo_discovery_2011, smercina_after_2022, luo_ic860_2022}, though the overall prevalence and impact of outflows on post-starburst evolution is not well understood.
In particular, it is unclear whether the primary effect of an outflow is to expel gas from a PSB \citep[e.g.][]{feruglio_quasar_2010, baron_evidence_2017}, or if the outflowing gas remains in the galaxy but disturbs the galaxy's gas reservoir, resulting in suppressed star-formation \citep[e.g.][]{alatalo_discovery_2011, luo_ic860_2022}.

We find evidence of an outflow ejecting molecular gas from the center in one of our PSBs, 9494-3701.
This is an edge-on disk galaxy with constant stellar and H$\alpha$ Radon profiles, though the H$\alpha$ PA is misaligned approximately 90$^\circ$ from the disk, as shown in Figure~\ref{fig:vmaps}.
The CO morphology in Figure~\ref{fig:overview_co} shows significant extraplanar gas also perpendicular to the disk which together appear to be a multiphase outflow.
However we do not see evidence of this outflow in the CO line profile (Figure~\ref{fig:overview_co}) or the position-velocity diagram (shown in Appendix~\ref{app:pv}), potentially because this extraplanar gas is too faint.
The low molecular gas fraction ($<$1\%) of the central post-starburst region may indicate that gas expulsion from the central region is contributing to quenching.
More investigation is needed to determine whether the outflows are powerful enough to eject the gas from the galaxy, and if the outflow itself is driving the quenching of the galaxy or if another physical process (such as an AGN) drives both the outflow and the central quenching.

It is unclear whether this is an AGN or starburst driven outflow.
The WHAN diagram in Figure~\ref{fig:overview_bpt} shows that 9494-3701 is dominated by retired spaxels and is thus ionized primarily by HOLMES.
However, given that the duty cycle of AGN is much shorter than the length of a post-starburst phase, it is possible that 9494-3701 previously hosted an AGN that has since shut off.
Alternatively, though the current central region star-formation surface density in the post-starburst region is very low (log $\Sigma_{SFR}$ = $-4.1 M_\odot$yr$^{-1}$kpc$^{-2}$), the recent starburst in this galaxy could have triggered the outflow and since quenched. 
Other galaxies in our sample with misaligned or disturbed gas kinematics, such as 8080-3704, 8655-1902, and 8941-3701 may also have outflows with masses too small to be detected with our CO data.
Detailed spectral fitting with high spectral resolution IFU data could provide strong outflow evidence in these galaxies.

Finally, we turn our attention to three galaxies without the signs of stellar or gas kinematic disturbance of the aforementioned galaxies: 8085-6104, 8086-3704, and 8982-6104. 
In principle, a number of processes could be driving their current post-starburst phases, including minor mergers or interactions, stellar bars, AGN feedback, or environmental processes like ram pressure stripping.
While none of these galaxies currently show obvious signs of a merger, 8086-3704 is our oldest PSB with a burst time of $3.0^{+0.4}_{-0.5}$ Gyr, so it is possible that the gas and stars have settled since a previous merger.

Alternatively, stellar bars could cause our observed centrally concentrated molecular gas reservoirs and star-formation suppression.
Bars can drive gas inflows, funneling gas inwards to the center of the galaxy \citep[e.g.][]{regan_inflow_1997, athanassoula_bar_2013, querejeta_torques_2016}.
Bars can also suppress star-formation in the disk \citep[e.g.][]{vera_bars_2016, kruk_barred_2018}, and may lead to increased star-formation in the gas-rich nuclear region \citep[e.g.][]{george_bar_2019}.

Only one of our galaxies has a strong bar, 8655-3701.
The bar is present in the SDSS image, and the position-velocity diagram of the CO emission shows the characteristic ``X" shape associated with a bar, shown in Appendix~\ref{app:pv}.
\citet{salim_spinning_2020} find that a rotating bar in NGC 7674, a Hickson compact group galaxy, likely drives turbulence leading to suppressed star-formation in the bar region.
While the nucleus of 8655-3701 is strongly star-forming, the ends of the bar are co-incident with post-starburst regions.
This quenching morphology is consistent with the rotation of the bar inducing star-formation suppressing turbulence as seen in NGC 7674, though these two galaxies reside in very different environments.
Higher resolution CO observations that can resolve the scale height of the gas for reliable velocity dispersion measurements are needed to confirm whether the bar is driving turbulence in this galaxy, and whether such turbulence is co-spatial with quenching regions.
It is also possible that the WHAN and BPT-classified AGN spaxels away from the center of this galaxy are from shocks which could be driven by the rotation of the bar, as has been observed in other strongly barred galaxies \citep[e.g.][]{athanassoula_dust_1992}.

While it is possible that other galaxies in our sample have weak bars undetected with the SDSS imaging, it is unclear whether a weak bar could drive star-formation suppression to the same extent as a strong bar, though bar length may be more important than bar strength \citep{fraser-mckelvie_spatially_2020, geron_bars_2021}.
The three galaxies without signs of kinematic disturbance (i.e. lacking asymmetrical stellar or H$\alpha$ Radon profiles and having kinematically aligned CO, stars, and H$\alpha$), 8085-6104, 8086-3704, and 8982-6104, all have inner bends in their stellar Radon profiles.
\citet{stark_radon_2018} found that stellar bars are the primary driver of inner bends in stellar Radon profiles, so a weak bar contributing to star-formation suppression in these galaxies is possible.

\subsection{Gas stabilization}

Figure~\ref{fig:sfe_fgas} shows that a substantial fraction of our post-starburst regions have significant gas reservoirs but suppressed star-formation.
This result is in good agreement with the global properties of PSBs from previous spatially unresolved studies \citep{rowlands_evolution_2015, french_clocking_2018}, but with our spatially resolved observations, we are considering only the post-starburst regions of our galaxies.
Thus, our extension of this result indicates that within a PSB, there is still a significant amount of gas in the quenching post-starburst region that is stabilized against star-formation.
We also see evidence of suppressed star-formation efficiency when comparing our galaxies to other star-forming samples along the Kennicutt-Schmidt relation, shown in Figure~\ref{fig:ks}.
The majority of our PSBs are below the relation measured by \citet{de_los_reyes_revisiting_2019}, with our cPSBs showing an especially large offset despite large molecular gas surface densities.
Though the source of this suppression is still unclear, shocks and turbulence driven by AGN feedback can plausibly suppress star-formation.

Shocks can regulate star-formation as their excess kinetic energy makes the gravitational collapse of molecular clouds needed for star-formation more difficult, thus decreasing star-formation efficiency.
The suppressive impact of shocks has been observed in a variety of studies, and in AGN host galaxies in particular \citep[e.g.][]{alatalo_strong_2014, guillard_turbulence_2015, aalto_excited_2015, alatalo_star_2015}.
Three of our galaxies have significant WHAN-classified star-forming centers with regions of off-center WHAN-classified AGN spaxels, consistent with shock ionization: 8083-12703, 8085-6104, and 8086-3704.
Three other galaxies have central AGN with separated off-center regions of AGN spaxels, consistent with AGN driven shocks: 8982-6104, 9088-9102, and 9194-3702.
The possible presence of shocks in 6/13 of our PSBs is suggestive, though from our data it is difficult to determine what the source of these shocks is.
Half of our shock candidates have AGN, indicating that AGN outflows could contribute to shocks.
In other galaxies, it is possible that the AGN has since shut off but the outflows are still creating shocks.
Mergers can also lead to shocks, which may be the case in 9088-9102 which is an on-going merger.
Finally, star-formation can drive outflows and shocks, as may be in the case of 8083-12703 and 8085-6104.
To unravel the impacts of shocks in our galaxies, we first must confirm these galaxies as shock hosts with UV and IR diagnostic lines \citep{kewley_understanding_2019}, while higher resolution CO observations will show whether shocked regions display greater star-formation suppression.

Turbulence from a recent merger, outflows, or other dynamic processes such as a stellar bar can also play a suppressive role in star-formation.
\citet{smercina_after_2018} find internal turbulent pressures in 3 PSBs $\sim$2 orders of magnitude greater than in typical star-forming galaxies, indicating that the extreme ISM conditions in PSBs are important for understanding the suppression of star-formation in these galaxies.
To maintain the turbulent pressures in these galaxies, a persistent source of turbulence is required, such as AGN-driven outflows or a changing gravitational potential caused by the resettling of the potential after a merger.
Thus, the tidal forces creating the observed asymmetries in the stellar velocity fields of our PSBs (see Section~\ref{sec:kinematics}) may lead to time-varying gravitational potentials which then drive turbulence in the gas, as observed in Arp 220, a dual nuclei merger \citep{scoville_arp_2017}.
Time invariant stellar dynamics in bulge-dominated galaxies can also drive turbulence in the gas and suppress star-formation through morphological quenching, resulting in a lack of star-forming dense gas \citep{martig_morphological_2009, martig_atlas3d_2013}.
However, morphological quenching impacts galaxies with low gas fractions ($<5$\%) and high stellar masses ($>3\times10^{10}$ M$_\odot$)
\citep{gensior_elephant_2021}, and none of our PSBs satisfy both of these requirements, making it unlikely that morphological quenching is significant in our sample.
Lastly, bar-driven quenching has been observed in a variety of studies as discussed in Section~\ref{sec:disc_tumult}
With kpc-scale resolution observations, we lack the spatial resolution to accurately measure the turbulent properties of the molecular gas.
Expanding high resolution CO observations to a larger sample of PSBs will shed light on the impact of a turbulent medium on star-formation.

The prevalence of AGN in 5/13 of our PSBs makes AGN feedback an attractive gas stabilization mechanism, though higher resolution observations are needed to test this possibility.
We classify 5 galaxies as potential AGN hosts with emission line ratios consistent with AGN in the WHAN diagram in the central region, as shown in Figure~\ref{fig:overview_bpt}.
For galaxies to quench in simulations, ``radio-mode'' AGN feedback is typically required to inject energy into gas reservoirs and stabilize them against gravitational collapse \citep{ciotti_feedback_2010}.
Studies of individual post-starbursts have drawn a connection between radio-mode AGN and star-formation suppression through more subtle means such as small outflows \citep[][]{alatalo_suppression_2015,luo_ic860_2022}.
However, the low spatial resolution of our CO and optical IFU data means we cannot resolve small-scale outflows and turbulence.
\citet{piotrowska_quenching_2021} find that radiatively inefficient AGN feedback drives turbulence and heats the ISM, leading to a decreased SFE.
Further, they find that this AGN feedback also heats the surrounding circumgalactic medium (CGM), thus preventing accretion of cold gas and reinvigoration of star-formation.
To confirm these galaxies as AGN hosts, high spatial resolution radio observations or deep X-ray data are needed, though we note that a lack of a radio or X-ray detection does not exclude the presence of an AGN.
Other galaxies in our sample may have recently hosted AGN that have shut off, as the duty cycle of AGN activity is much shorter than the PSB phase \citep{pawlik_origins_2018}.
Thus, the lack of current evidence for AGN activity does not eliminate the possibility that AGN feedback is important in regulating star-formation in our entire sample, as found in \citet{piotrowska_quenching_2021}.
High resolution, multiwavelength observations are needed to determine the galactic-scale processes suppressing star-formation and to study the impacts these mechanisms have on the cloud-scale ISM conditions governing star-formation.

\subsection{Central and ring post-starbursts}

Overall, it is difficult to distinguish the quenching mechanisms active in ring and central PSBs with our data and small sample.
\citet{chen_post-starburst_2019} find that cPSBs have a more dramatic history, while rPSBs with star-forming centers may be quenching due to gas supply cutoff in the outskirts.
Morphologically, cPSBs and rPSBs have similar mean core fractions (67\% and 65\% respectively, see Figure~\ref{fig:fcore_psf}).
Both cPSBs and rPSBs show asymmetrical stellar Radon profiles (3/7 and 2/4 respectively), and both samples include galaxies with kinematic misalignments (6/9 cPSBs and 1/4 rPSBs).
Both central and ring post-starburst regions show evidence of star-formation suppression, though to a lesser degree for the ring post-starburst regions than some central regions, as shown in Figures~\ref{fig:ks} and \ref{fig:fgas_sfh}.
Finally, both samples have a significant portion of central AGN ionization (3/9 for cPSBs and 2/4 for rPSBs).
The two rPSBs without central AGN spaxels instead appear to be ionized by shocks in the outskirts of the galaxy with star-formation dominating the central ionization, similar to 8086-3704, a cPSB.
Overall, it appears that both cPSBs and rPSBs often have undergone a recent merger, given the high rate of stellar asymmetries and centrally concentrated molecular gas in both samples. 

The natural question that arises is why do some mergers appear to lead to a central PSB whereas others lead to a ring PSB?
Mergers can take many different forms with varying stellar mass ratios, gas fractions, and orientations all playing significant roles.
More theoretical work is needed to determine the qualities that lead a merger remnant to a cPSB versus an rPSB phase.
One area where we observe a difference between our cPSBs and rPSBs is in the gas kinematics.
In Figure~\ref{fig:radon}, we see that 4/8 cPSBs have asymmetric H$\alpha$ Radon profiles, while 0/4 rPSBs do.
Similarly, looking at the CO velocity maps in Figure~\ref{fig:vmaps}, 4/8 CO-detected cPSBs lack significant CO rotation, whereas all 4 rPSBs have strong CO rotation (though it is misaligned in one case).
While the small sample of our rPSBs makes any conclusions tentative, it appears that the gas in rPSBs is less disturbed than in cPSBs, potentially from a less disruptive merger or from a faster resettling of the gas.
It is also possible that the highly disturbed gas in cPSBs is driven by a process more commonly active in cPSBs, such as outflows. 

Our conclusion that mergers are important in the formation of both cPSBs and rPSBs is consistent with \citet{chen_post-starburst_2019}.
With the addition of our spatially resolved CO observations, we find that the gas appears to be less disturbed in rPSBs, showing stronger rotation and more symmetrical velocity maps.
\citet{chen_post-starburst_2019} differentiate between two types of rPSBs, those that are experiencing a global shutdown of star-formation, and those with star-forming centers.
They conclude that the former group may be quenching from a lack of gas inflow, while the latter likely have experienced either ram pressure stripping or a merger/interaction.
\citet{owers_hdsg_2019} also identify galaxies with central star-formation and post-starburst signatures in the outskirts, though they find these galaxies almost exclusively in a cluster context, and conclude that the outer post-starburst regions are driven by ram-pressure stripping.
Our rPSBs similarly have star-forming centers and outer post-starburst regions.
However, the post-starburst regions in our rPSBs have similar gas fractions to the central post-starburst regions and ALMaQUEST star-forming galaxies, indicating that a lack of molecular gas is not driving quenching.
The primary difference \citet{chen_post-starburst_2019} find between their cPSBs and rPSBs are in their star-formation histories and rotational support, where cPSBs have a more mixed-age stellar population throughout the galaxy, and also have less rotational support, both of which indicate a recent dramatic event such as a merger.
Our results show stellar disturbances in both cPSBs and rPSBs, indicating a common merger origin, but the molecular and ionized gas in cPSBs appears to be more disturbed than in rPSBs, hinting at the importance of gas disruption through processes such as outflows.

It is tempting to point towards rPSBs as outside-in quenching systems and cPSBs as inside-out quenching systems, but this simplistic picture has a number of flaws.
It is unclear whether rPSBs can truly be considered ``quenching" systems, as their evolution is unknown: will the outskirts of these galaxies eventually undergo re-invigoration of their star-formation, or is an rPSB phase the first step towards outside-in quenching?
To answer this question, a larger sample of rPSBs is needed with a range of post-burst ages, as well as an improved theoretical understanding of how rPSBs form and evolve.
IFS studies of cPSBs have enabled measurements of radial gradients in stellar populations.
\citet{chen_post-starburst_2019} find that the outskirts of cPSBs show weaker post-starburst signatures than the central regions, indicating that these regions either never underwent a starburst or that the burst was weaker or older.
\citet{zheng_comparison_2020} use binary merger simulations to measure radial gradients in cPSBs and find that the starburst peaks at the same time throughout the galaxy, but is more prolonged and stronger in the center, leading to the observed radial gradients in \citet{chen_post-starburst_2019}.
In our sample, one of our cPSBs, 8941-3701, has a central post-starburst region and the outskirts of the galaxy are quiescent but not post-starburst, which is consistent with outside-in quenching.
More detailed spatially resolved spectral fitting on a larger sample will show whether this qualitative outside-in behavior is typical.

\section{Conclusions} \label{sec:conclusions}
In this work we present ALMA CO(1-0) observations of 13 post-starburst galaxies with accompanying MaNGA optical IFS data, 9 of which have central post-starburst regions and 4 with ring outskirt post-starburst regions.
With these data, we study the molecular gas morphology, kinematics, and star-formation properties of these galaxies to work towards understanding the causes and mechanisms of star-formation quenching in these galaxies.
We list our primary conclusions below:
\begin{enumerate}
    \item \textit{The majority of our PSBs show evidence of a recent merger.}
    7/13 PSBs have asymmetric or non-rotating stellar velocity fields, consistent with tidal forces from a merger or interaction disrupting the stellar kinematics.
    The remaining six galaxies lack kinematic evidence of a recent merger.
    While mergers resulting in symmetric stellar disturbances (such as a disk warp) and minor mergers are not ruled out, other processes such as outflows and weak stellar bars may play a role in the quenching of these galaxies, as we see in outflow host 9494-3701.
    
    \item \textit{Significant molecular gas reservoirs in post-starburst regions are stabilized against star-formation.}
    Within the post-starburst regions of our galaxies, we find typical molecular gas fractions but low star-formation efficiencies in nearly half of our sample.
    PSBs then do not typically quench due to a lack of molecular gas but because the existing gas is not forming stars.
    AGN feedback may drive this stabilization, with 5/13 galaxies hosting central AGN-like emission.
    
    \item \textit{Central and ring post-starburst regions have similar properties, though gas in rPSBs may be less disturbed.}
    The post-starburst regions of our central and ring PSB regions have similar star-formation and molecular gas properties.
    Both samples have approximately half of galaxies with evidence of mergers (3/7 cPSBs and 2/4 rPSBs).
    However, both the ionized and molecular gas in rPSBs appears to be less disturbed than that of the cPSBs, which may hint at different merger conditions or another process primarily impacting cPSBs, such as outflows. 

\end{enumerate}

\clearpage

\begin{acknowledgments}
We thank the anonymous referee for helpful comments that improved this work.

{JO, YL, KA, KR, and ES have been have been partially funded by Space Telescope Science Institute Director's Discretionary Research Fund grants D0101.90241, D0101.90276, D0101.90262, D0101.90281, D0101.90296, \textit{HST} grants GO-14715.021, GO-14649.015, and \textit{Chandra} grant GO7-18096A.
JO acknowledges support from NRAO under grant No. SOSPA7-027.
LL thanks the support from the Academia Sinica under the Career Development Award CDA107-M03 and the Ministry of Science \& Technology of Taiwan under the grant MOST 111-2112-M-001-044.
HAP acknowledges support by the National Science and Technology Council of Taiwan under grant 110-2112-M-032-020-MY3.
WB acknowledges support from the Science and Technology Facilities Council (STFC).
The Cosmic Dawn Center is funded by the Danish National Research Foundation under grant No. 140. 
RAR acknowledges the support from Conselho Nacional de Desenvolvimento Cient\'ifico e Tecnol\'ogico  and Funda\c c\~ao de Amparo \`a pesquisa do Estado do Rio Grande do Sul.
TAD acknowledges support from the UK Science and Technology Facilities Council through grants ST/S00033X/1 and ST/W000830/1.
FvdV was supported by a Royal Society University Research Fellowship.}

{The National Radio Astronomy Observatory is a facility of the National Science Foundation operated 
under cooperative agreement by Associated Universities, Inc.

This paper makes use of the following ALMA data: ADS/JAO.ALMA\#2019.1.01136.S. ALMA is a partnership of ESO (representing its member states), NSF (USA) and NINS (Japan), together with NRC (Canada), MOST and ASIAA (Taiwan), and KASI (Republic of Korea), in cooperation with the Republic of Chile. The Joint ALMA Observatory is operated by ESO, AUI/NRAO and NAOJ.}

{Funding for the Sloan Digital Sky 
Survey IV has been provided by the 
Alfred P. Sloan Foundation, the U.S. 
Department of Energy Office of 
Science, and the Participating 
Institutions. 

SDSS-IV acknowledges support and 
resources from the Center for High 
Performance Computing  at the 
University of Utah. The SDSS 
website is www.sdss.org.}
\end{acknowledgments}

\begin{acknowledgments}
SDSS-IV is managed by the 
Astrophysical Research Consortium 
for the Participating Institutions 
of the SDSS Collaboration including 
the Brazilian Participation Group, 
the Carnegie Institution for Science, 
Carnegie Mellon University, Center for 
Astrophysics | Harvard \& 
Smithsonian, the Chilean Participation 
Group, the French Participation Group, 
Instituto de Astrof\'isica de 
Canarias, The Johns Hopkins 
University, Kavli Institute for the 
Physics and Mathematics of the 
Universe (IPMU) / University of 
Tokyo, the Korean Participation Group, 
Lawrence Berkeley National Laboratory,
Leibniz Institut f\"ur Astrophysik
Potsdam (AIP), Max-Planck-Institut 
f\"ur Astronomie (MPIA Heidelberg), 
Max-Planck-Institut f\"ur 
Astrophysik (MPA Garching), 
Max-Planck-Institut f\"ur 
Extraterrestrische Physik (MPE), 
National Astronomical Observatories of 
China, New Mexico State University, 
New York University, University of 
Notre Dame, Observat\'ario 
Nacional / MCTI, The Ohio State 
University, Pennsylvania State 
University, Shanghai 
Astronomical Observatory, United 
Kingdom Participation Group, 
Universidad Nacional Aut\'onoma 
de M\'exico, University of Arizona, 
University of Colorado Boulder, 
University of Oxford, University of 
Portsmouth, University of Utah, 
University of Virginia, University 
of Washington, University of 
Wisconsin, Vanderbilt University, 
and Yale University.
\end{acknowledgments}

\vspace{5mm}
\facilities{ALMA, Sloan}

\software{This research has made use of the CIRADA cutout service at URL cutouts.cirada.ca, operated by the Canadian Initiative for Radio Astronomy Data Analysis (CIRADA). CIRADA is funded by a grant from the Canada Foundation for Innovation 2017 Innovation Fund (Project 35999), as well as by the Provinces of Ontario, British Columbia, Alberta, Manitoba and Quebec, in collaboration with the National Research Council of Canada, the US National Radio Astronomy Observatory and Australia’s Commonwealth Scientific and Industrial Research Organisation.
We use the \texttt{NADA} package in R to compute the Kendall's $\tau$ coefficients and accompanying p-values.
Astropy \citep{astropy_2013, astropy_2018}, CASA \citep[V6.4.3;][]{mcmullin_casa_2007}, celerite2 \citep{celerite1, celerite2}, mangadap \citep{westfall_dap_2019}, Marvin \citep{marvin_paper_2019}, matplotlib \citep{matplotlib_2007}, numpy \citep{numpy_2020}, PaFit \citep{krajnovic_kinemetry_2006}, statmorph \citep{rodriguez-gomez_optical_2019}.
}

\bibliography{psb_bib.bib}
\bibliographystyle{aasjournal}

\appendix

\section{PCA classification boundaries} \label{app:pca}

\begin{figure*}[hbt!]
    \centering
    \includegraphics[width=\textwidth]{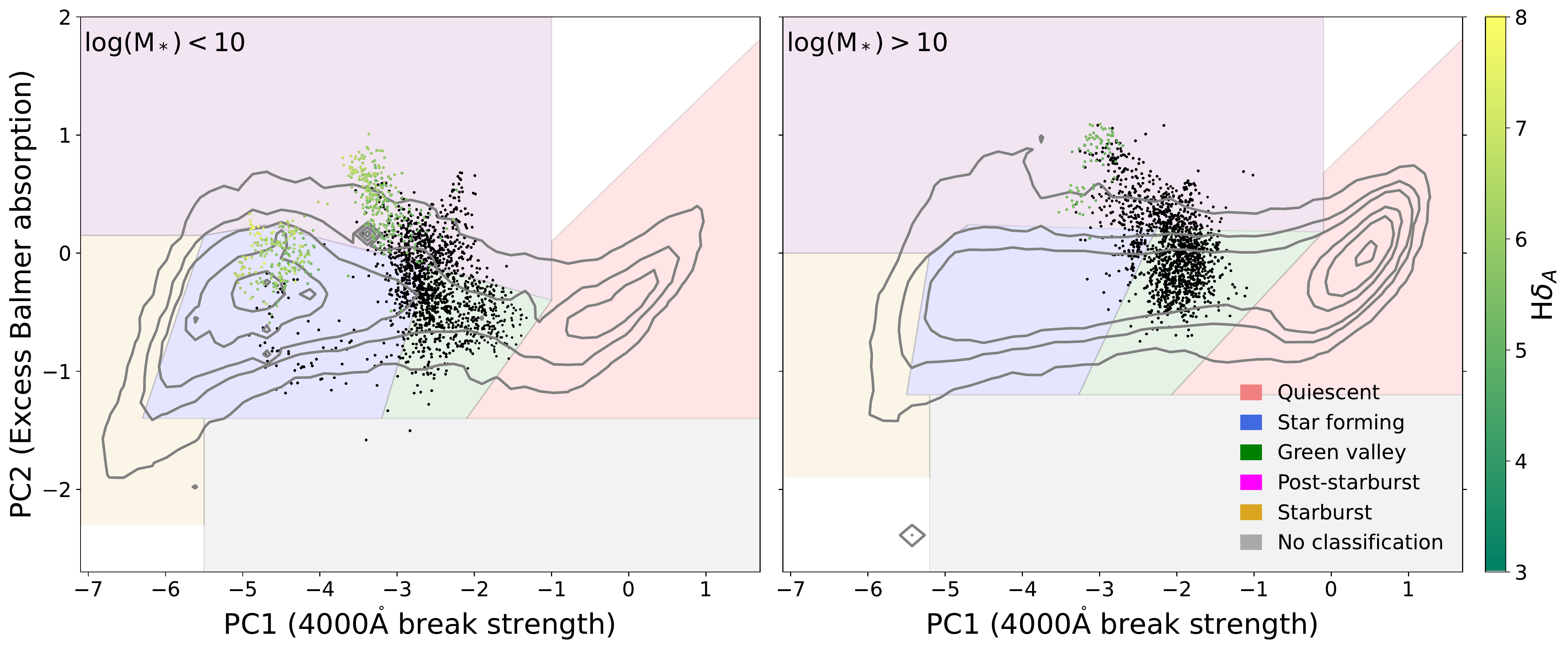}
    \caption{PCA classification boundaries in PC1 (D$_n$4000) and PC2 (excess Balmer absorption).
    The shaded regions show the boundaries for quiescent (red), star-forming (blue), green valley (green), post-starburst (purple), and starburst (yellow) classifications.
    The gray contours show the parent DR15 MaNGA sample, with levels corresponding to 1\%, 5\%, 30\%, 50\%, and 90\% of the largest bin.
    The black and colored points show all spaxels from galaxies in the \citet{chen_post-starburst_2019} and PCA combined sample.
    The colored points are spaxels within those galaxies that are classified as post-starburst with the \citet{chen_post-starburst_2019} selection method, and are color coded according to their H$\delta_A$.}
    \label{fig:pca_class}
\end{figure*}

We show the classification boundaries for the PCA analysis in Figure~\ref{fig:pca_class}.
The boundaries differ for low and high mass spaxels (defined as above and below $M_* = 10^{10}$ M$_\odot$) as PC1 and PC2 values have a weak stellar mass dependence.
We modified the original boundaries from \citet{rowlands_sdss-iv_2018} visually to optimize the agreement between the PCA method and the \citet{chen_post-starburst_2019} selection method while minimizing contamination. 
We plot all spaxels from the final \citet{chen_post-starburst_2019} and PCA combined sample in Figure~\ref{fig:pca_class}, and color spaxels identified as post-starburst by \citet{chen_post-starburst_2019} in green according to their H$\delta_A$ value.
In the high mass case, all \citet{chen_post-starburst_2019} PSB-classified spaxels are also PSB-classified with the PCA method, though for spaxels in low stellar mass galaxies there is overlap with the star-forming classification region.

Different post-starburst selection methods result in samples with different properties including post-burst ages and AGN or shock presence, and no single selection will include all galaxies considered to be post-starburst.

\section{8939-3703: a star-forming interloper}\label{app:8939}

As mentioned in Section~\ref{sec:sample}, our spectral fitting results of 8939-3703 indicate that this galaxy is a star-forming interloper.
We plot the star-formation history of this galaxy in Figure~\ref{fig:SFH_all}, where we see relatively constant star-formation over the last $\sim$3 Gyr. 
Since the residuals are in line with Gaussian random noise and the fitted Gaussian process term is well behaved, the optical continuum for this galaxy is well fit.
The derived SFH indicates that there is little evidence in the continuum that a recent starburst nor rapid quenching occurred in the PSB spaxels.

\begin{figure}
    \centering
    \includegraphics[width=0.8\textwidth]{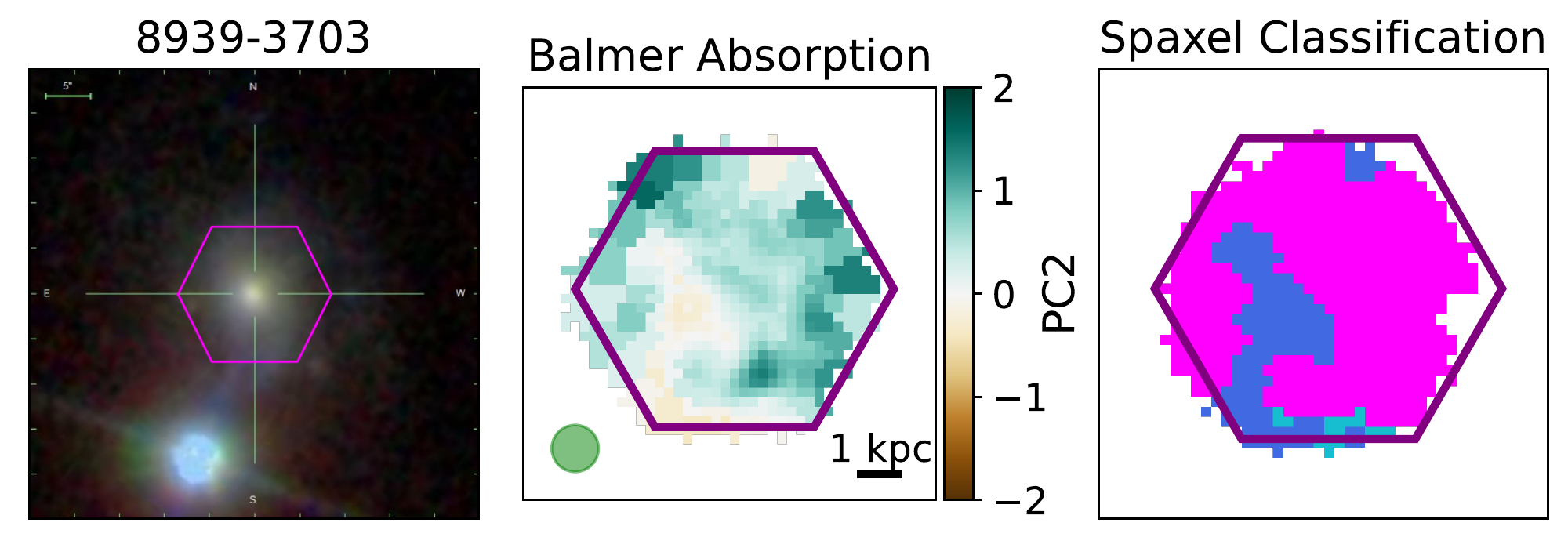}
    \caption{PCA classification results for 8939-3703. Left: the SDSS 3-color image, with a clear foreground star. Center: the excess Balmer absorption of each spaxel. Teal regions correspond to greater Balmer absorption, and to PSB regions. Right: the PCA classification from \citet{rowlands_sdss-iv_2018}, with post-starburst spaxels in purple, star-forming spaxels in blue, and starburst spaxels in teal. The green circle shows the MaNGA 2.5\arcsec\, average seeing. This galaxy is clearly classified as a PSB.}
    \label{fig:8939}
\end{figure}

This galaxy is selected as a PSB using the PCA method of \citet{rowlands_sdss-iv_2018}, though we note it is not selected in the sample of \citet{chen_post-starburst_2019}.
Figure~\ref{fig:8939} shows the SDSS 3-color image, the PCA spaxel classifications, and the excess Balmer absorption of 8939-3703.
The high values of excess Balmer absorption mean that many spaxels in this galaxy are not close to the PSB classification boundary in D$_n$4000 and Balmer absorption space (principal components 1 and 2 respectively in our PCA analysis).
In the optical image we do see that a foreground star is close to the galaxy.
However, there is no visible diffraction spike in the MaNGA data and all the quality flags indicate this star did not contaminate the data.

We do not detect any CO(1-0) emission from this galaxy with ALMA.
We measure an upper limit molecular gas mass of $\log M_{H_2}/M_\odot < 7.7$.

\section[Different values of alphaCO]{Different values of $\alpha_{CO}$}\label{app:alphaCO}

Observationally, \citet{smercina_after_2018} find that an $\alpha_{CO}$ of 4 M$_\odot$ (K km s$^{-1}$ pc$^2$)$^{-1}$ was generally in agreement with molecular gas masses derived from fitting multiple H$_2$ rotational emission lines for a sample of E+A galaxies.
However, their requirement of W(H$\alpha$) $<$ 3 \AA\, for their sample would likely exclude 8/13 galaxies in our sample, so we investigate the impact of a varying $\alpha_{CO}$ on our results.

The unique ISM conditions in PSBs makes determining the proper conversion from CO line flux to molecular gas mass, i.e. $\alpha_{CO}$ difficult.
Primary factors that can alter $\alpha_{CO}$ include metallicity and the galactic environment - often including starbursting galaxies and mergers \citep[e.g.][]{bolatto_co--h2_2013}.
\citet{narayanan_general_2012} combine these two impacts in a single prescription for $\alpha_{CO}$:

\begin{equation} \label{eq:alphaCO_var}
    \alpha_{CO} = \frac{\textrm{min}[6.3, 10.7 \times \langle W_{CO} \rangle^{-0.32}}{Z'^{0.65}}
\end{equation}

Where $\langle W_{CO}\rangle$ is the CO surface brightness measured in K km/s, and $Z'$ is the gas-phase metallicity normalized by the Solar metallicity.
While we directly measure the CO surface brightness, determining the gas-phase metallicity is difficult in the non star-forming spaxels of our PSBs, though work is being done to extend metallicity relations to LI(N)ER regions \citep{kumari_extension_2021}.
\citet{goto_abundances_2007} find that the metallicities of a sample of 451 E+A are consistent with solar metallicities.
Given that AGN appear to be present in about 4 of our galaxies (see Section~\ref{sec:emlines}) and would not be included in the E+A sample of \citet{goto_abundances_2007}, we also consider that the typical metallicity of AGN hosts can range from near-solar values to up to a few times solar metallicity \citep[e.g.][]{hamann_metallicities_2002}.
We conservatively assign a metallicity of two times solar (i.e. $Z' = 2$) for our sample and compute new $\alpha_{CO}$ values for our post-starburst regions.
The mean $\alpha_{CO}$ of our post-starburst regions using Equation~\ref{eq:alphaCO_var} is 1.9 M$_\odot$ (K km s$^{-1}$ pc$^2$)$^{-1}$, which is within a reasonable uncertainty of a Milky-Way $\alpha_{CO}$ of 4.35 M$_\odot$ (K km s$^{-1}$ pc$^2$)$^{-1}$ of a factor of a few.
Alternatively, if our PSBs were to have a sub-solar metallicity of $Z' = 0.5$, we find a mean $\alpha_{CO} = 4.75$ M$_\odot$ (K km s$^{-1}$ pc$^2$)$^{-1}$, thus slightly increasing our measured gas masses, but well within our uncertainties.
A sub-solar metallicity could be achieved from pristine gas inflow. 

\begin{figure}
    \centering
    \includegraphics[width=0.4\textwidth]{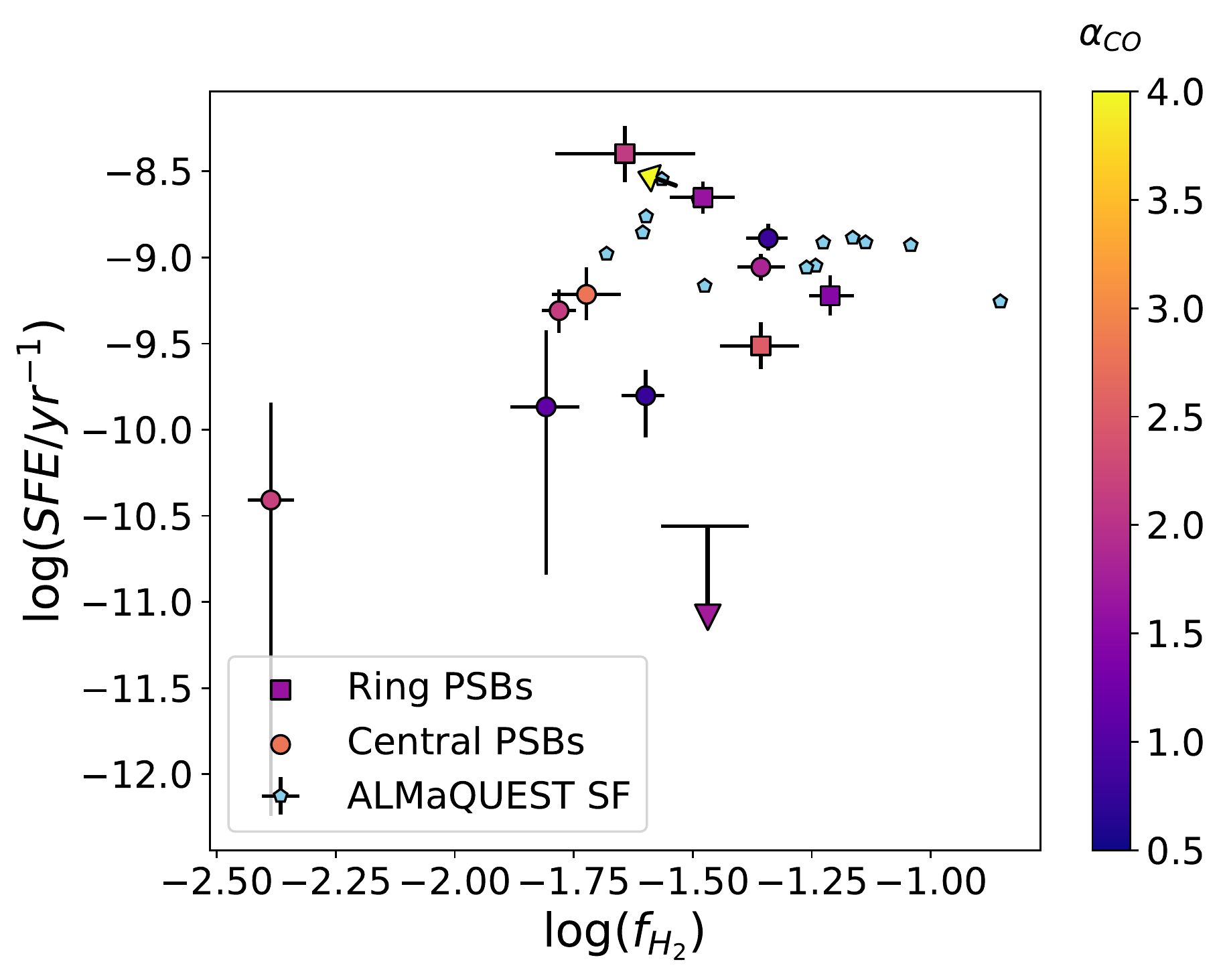}
    \caption{Star-formation efficiencies and molecular gas fractions computed with a varying $\alpha_{CO}$ from \citet{narayanan_general_2012} for post-starburst regions in our sample, with $Z' = 2$. Post-starburst regions of cPSBs are plotted as circles, and post-starburst regions of rPSBs are plotted as squares. These markers are colored by their computed $\alpha_{CO}$ value. We also plot ALMaQUEST star-forming galaxies, with a Milky-Way $\alpha_{CO}$ of 4.35 M$_\odot$ (K km s$^{-1}$ pc$^2$)$^{-1}$. }
    \label{fig:sfe_fgas_alt}
\end{figure}

In Figure~\ref{fig:sfe_fgas_alt}, we plot the gas fraction and star-formation efficiency of our post-starburst regions (as in Figure~\ref{fig:sfe_fgas}) but with the variable $\alpha_{CO}$ computed with Equation~\ref{eq:alphaCO_var}, assuming $Z' = 2$.
Though the gas fractions are lower and star-formation efficiencies higher than with the Milky Way $\alpha_{CO}$ in Figure~\ref{fig:sfe_fgas}, our primary conclusion that the many of our post-starburst regions have similar molecular gas fractions to the star-forming galaxies with lower star-formation efficiencies still holds.

\section{Fitting star-formation histories}\label{app:SFH}

Table~\ref{tab:priors} lists the 18 parameters in the final SFH model, with 15 free to vary with priors.
The parameters are described in Section~\ref{sec:fitting}. 

\begin{table*}
\centering
    \begin{tabular}{l|l|l|l|l}
    Type          & Parameter                     & Form                & Min       & Max     \\ \hline
    SFH           & $\log_{10}M_*/M_\odot$        & Uniform             & 6         & 13      \\
                  & $t_{\rm{form}}$ (Gyr)         & Uniform             & 4         & 14      \\
                  & $\tau_e$ (Gyr)                & Uniform             & 0.3       & 10      \\
                  & $t_{\rm{burst}}$ (Gyr)        & Uniform             & 0         & 4       \\
                  & $\alpha$                      & $\log_{10}$ Uniform & 0.01      & 1000    \\
                  & $\beta$                       & Fixed = 250         & -         & -       \\
                  & $f_{\rm{burst}}$              & Uniform             & 0         & 1       \\
    Metallicity   & $Z_{\rm{old}}/Z_\odot$        & $\log_{10}$ Uniform & 0.01      & 2.5     \\
                  & $Z_{\rm{burst}}/Z_\odot$      & $\log_{10}$ Uniform & 0.01      & 2.5     \\
    Dust          & $A_V$                         & Uniform             & 0         & 2       \\
                  & birthcloud factor $\eta$      & Uniform             & 1         & 5       \\
    GP noise      & uncorrelated amplitude $s$    & $\log_{10}$ Uniform & 0.1       & 10      \\
                  & correlated amplitude $\sigma$ & $\log_{10}$ Uniform & $10^{-4}$ & 1       \\
                  & period/length scale $\rho$    & $\log_{10}$ Uniform & 0.04      & 1       \\
                  & dampening quality factor $Q$  & Fixed = 0.49        & -         & -       \\
    Miscellaneous & redshift                      & Uniform             & 0.8 $z$   & 1.2 $z$ \\
                  & $t_{\rm{birth cloud}}$ (Gyr)  & Fixed = 0.01        & -         & -       \\
                  & $\sigma_{\rm{disp}}$ (km/s)   & $\log_{10}$ Uniform & 40        & 4000   
    \end{tabular}
    \caption{Model priors used for spectral fitting of stacked PSB-only spaxels. The star-formation history parameter symbols are described in equations \ref{eq:psb_wild2020} to \ref{eq:dpl}; the Gaussian process parameter symbols are described in \cite{celerite1,celerite2} and Leung et al. (in prep.). Other symbols have their usual meanings. A $\log_{10}$ uniform prior form indicates a flat prior in $\log(X) \sim U(\log(min), \log(max))$. Redshift is given a uniform prior from 80\% to 120\% of the target's catalog value.}
\label{tab:priors}
\end{table*}

\begin{figure*}
    \centering
    \includegraphics[width=0.9\textwidth]{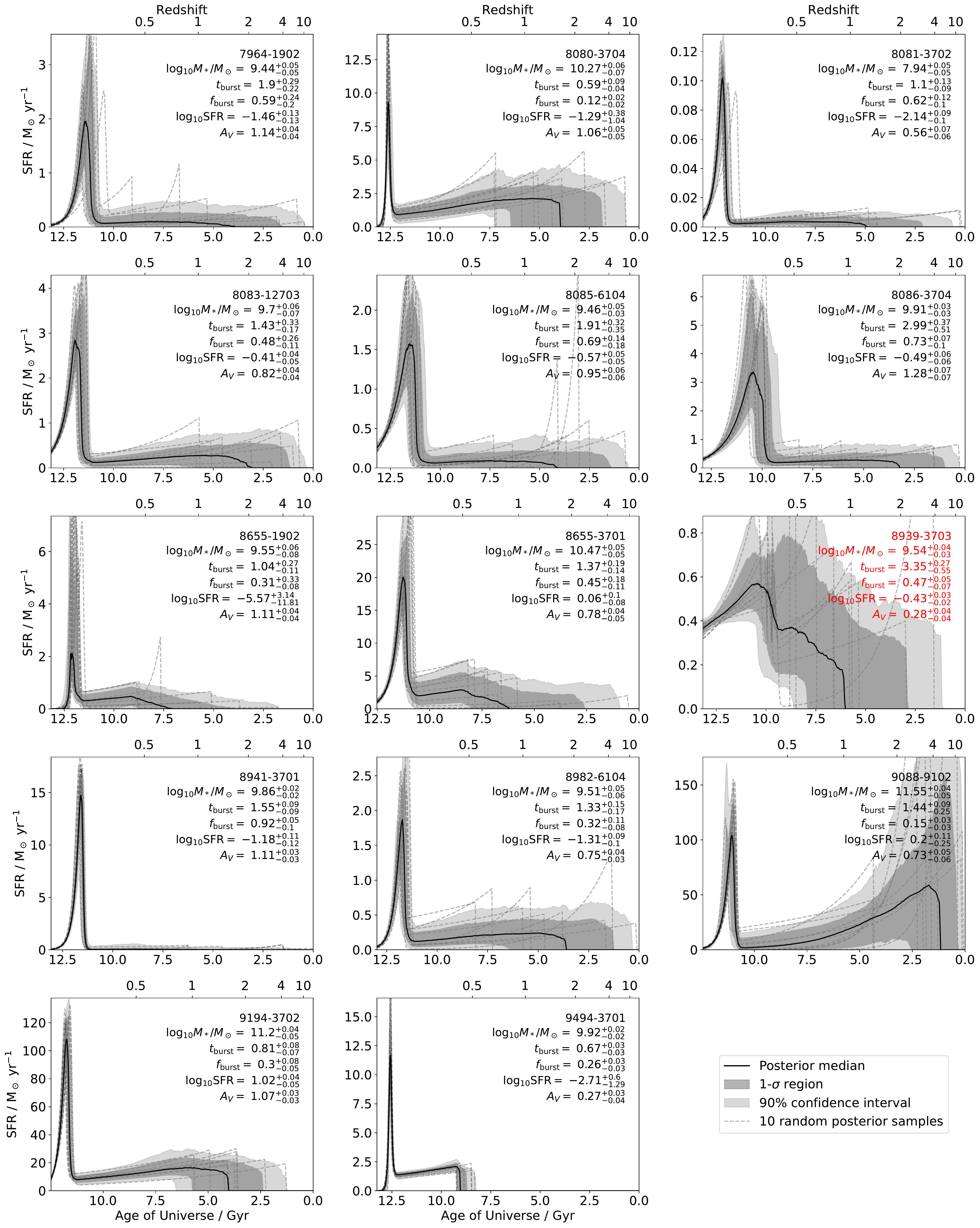}
    \caption{A montage of fitted star-formation histories (SFH) through our spectral fitting methods (Section~\ref{sec:fitting}). In each panel, the MaNGA Plate-IFU, along with several key fitted properties (of only the PSB regions) are listed towards the top right. Their symbols correspond to those in Tables \ref{tab:priors} and \ref{tab:region}. 8939-3703, highlighted with red text, is found to be a star-forming interloper (see Appendix~\ref{app:8939}). The bottom x-axis traces age of the universe, with the corresponding redshift scale displayed on the top edge, while the y-axis shows star-formation rates. The star-formation histories grow progressively more uncertain with greater lookback time.}
    \label{fig:SFH_all}
\end{figure*}

Figure~\ref{fig:SFH_all} shows our derived SFH's for each galaxy in our sample.
The spectral fitting methods used are described in Section~\ref{sec:fitting}.
SFH parameters such as the time since burst, burst mass fraction, etc are listed for each galaxy in Table~\ref{tab:region}.

\section{Position-velocity diagrams} \label{app:pv}

\begin{figure*}[hbt!]
    \centering
    \includegraphics[width=0.9\textwidth]{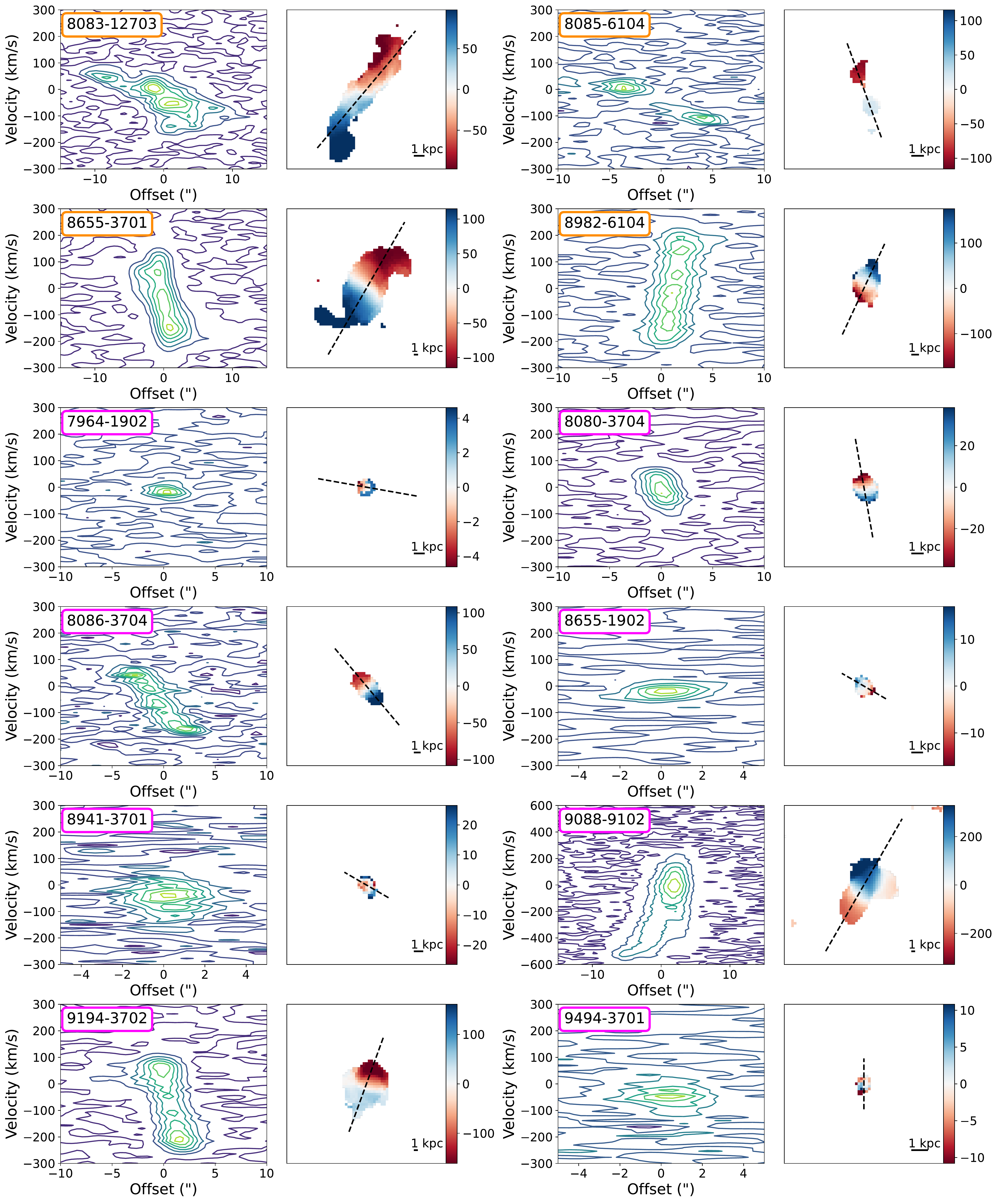}
    \caption{CO position-velocity diagrams for each galaxy with a CO detection. The right panel shows the CO velocity field with a dashed line showing the axis where the position-velocity diagram on the left is extracted from. The colorbar is in units of km/s.}
    \label{fig:pv_diagrams}
\end{figure*}

In Figure~\ref{fig:pv_diagrams}, we plot the CO position-velocity (P-V) diagram for each galaxy with a CO detection.
We select the axis along which we extract the P-V diagram by hand with a width of 1\arcsec.
We see that 6 galaxies have rotating CO emission, though these galaxies do not have a flattening of their rotation curves at larger radii, possibly as a result of the centrally concentrated gas morphologies. 
One galaxy has a bar, 8655-3701, as shown by the ``X" shape in the P-V diagram.
This galaxy also has an obvious stellar bar in the optical image.
Finally, the P-V diagram of 8083-12703 shows a complex kinematic structure with a radius-dependent kinematic position angle, which may be indicative of some disturbance such as a minor merger or interaction.

\section{3mm and radio continuum detections} \label{app:radio}

Radio and millimeter continuum detections can be associated with an AGN or a starburst \citep[e.g.][]{best_radio_2005}. 
We detect 3mm continuum emission to 5$\sigma$ in 7/13 of our PSBs.
For each galaxy, this emission is unresolved and in the center of the galaxy except for 9088-9102, which has extended continuum emission in the center, and 9494-3701, which has continuum emission offset from the center.

\begin{figure*}[hbt!]
    \centering
    \includegraphics[width=\textwidth]{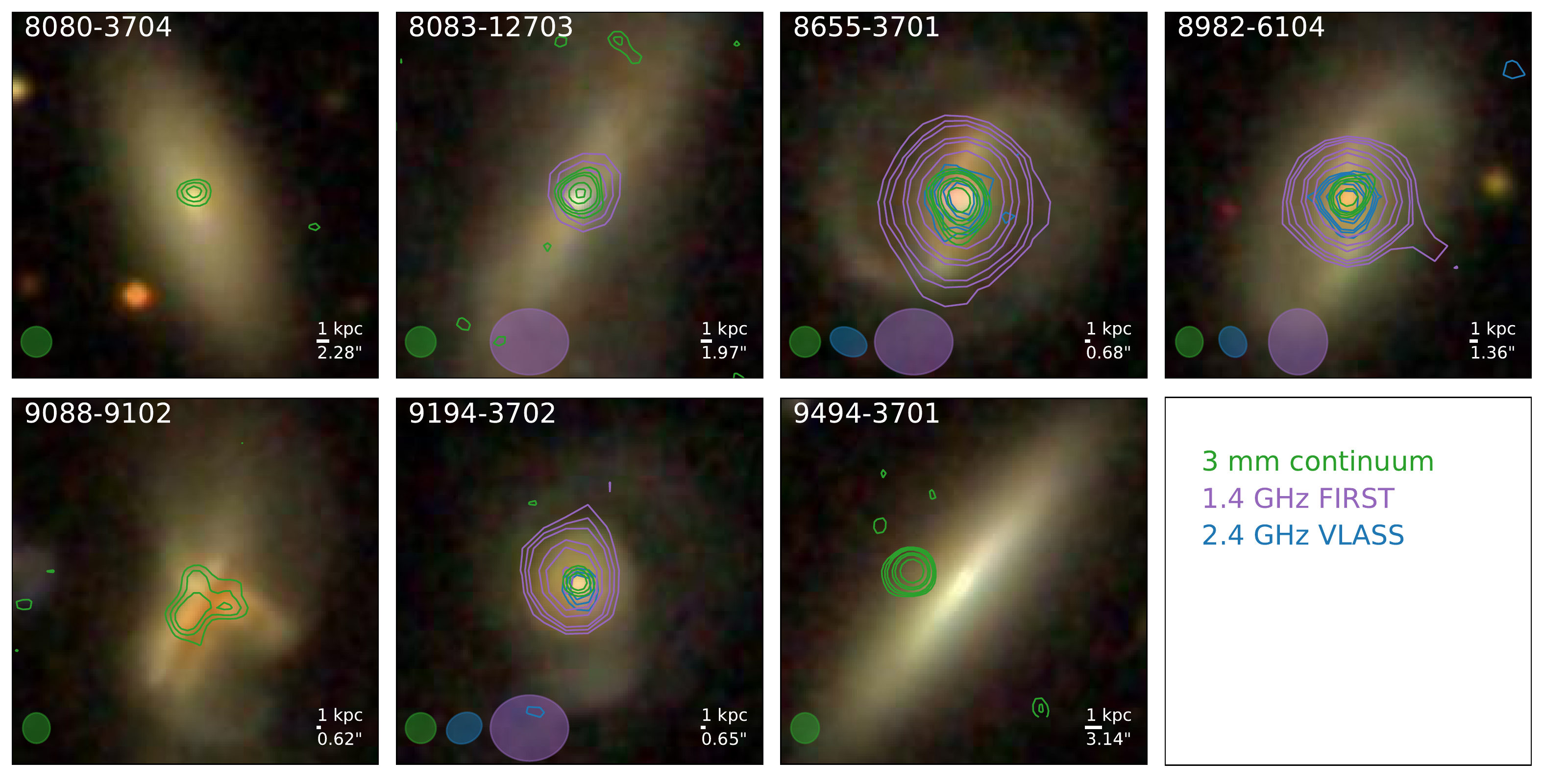}
    \caption{Millimeter and radio emission contours overlaid on SDSS 3-color images for galaxies with 5$\sigma$ millimeter or radio continuum detections. Green contours show the 3 mm continuum emission, purple contours show the FIRST 1.4 GHz emission, and blue contours show the VLASS 2.4 GHz emission. The ellipses in the lower left of each panel show the beam sizes with the same colors as the contours. Cutout images from the VLASS survey were accessed with \texttt{astroquery} \citep{ginsburg_astroquery_2019}.}
    \label{fig:overview_cont}
\end{figure*}

We measure radio fluxes of our galaxies from cutouts from the VLA Sky Survey \citep[VLASS,][]{lacy_karl_2020} and Faint Images of the Radio Sky at Twenty centimeters survey \citep[FIRST,][]{becker_vlas_1994} at the optical source locations of our galaxies.
We first fit the cutouts with a 2D Gaussian, and measure fluxes within an elliptical aperture corresponding to the 3$\sigma$ Gaussian fit size. 
We measure 4 galaxies with S/N $>$ 3 from the FIRST survey, and 3 from VLASS.
We show the 3 mm continuum contours for galaxies with a 5$\sigma$ continuum millimeter or radio detection, overlaid on the 3-color SDSS image in Figure~\ref{fig:overview_cont}.
For galaxies with radio continuum detections, the emission is co-incident with the millimeter continuum emission.

For the majority of 3mm continuum detected galaxies, AGN likely contribute to the 3mm emission, though star-formation may also contribute significantly as a dust heating source and through free-free emission.
Of the 7 galaxies with 3mm continuum emission, 5 have central emission line ratios consistent with some AGN contribution (see Section~\ref{sec:emlines}).
8083-12703 is an rPSB with 3mm emission and no evidence of a central AGN but does have a star-forming center, indicating that the central, compact 3mm emission could be from a central starburst.
The extended nature of the 3mm emission in 9088-9102 also points towards a significant contribution from star-formation, as we would expect continuum emission heated by a central AGN to be highly compact.
Lastly, the offset of the continuum emission of 9494-3701 and optical center of the galaxy indicates that this emission may be from a background source rather than from within the galaxy itself.
None of the 12 ALMaQUEST star-forming galaxies have 3 mm continuum detections.

\section{Half-light radii}

\begin{figure}[hbt!]
    \centering
    \includegraphics[width=0.4\textwidth]{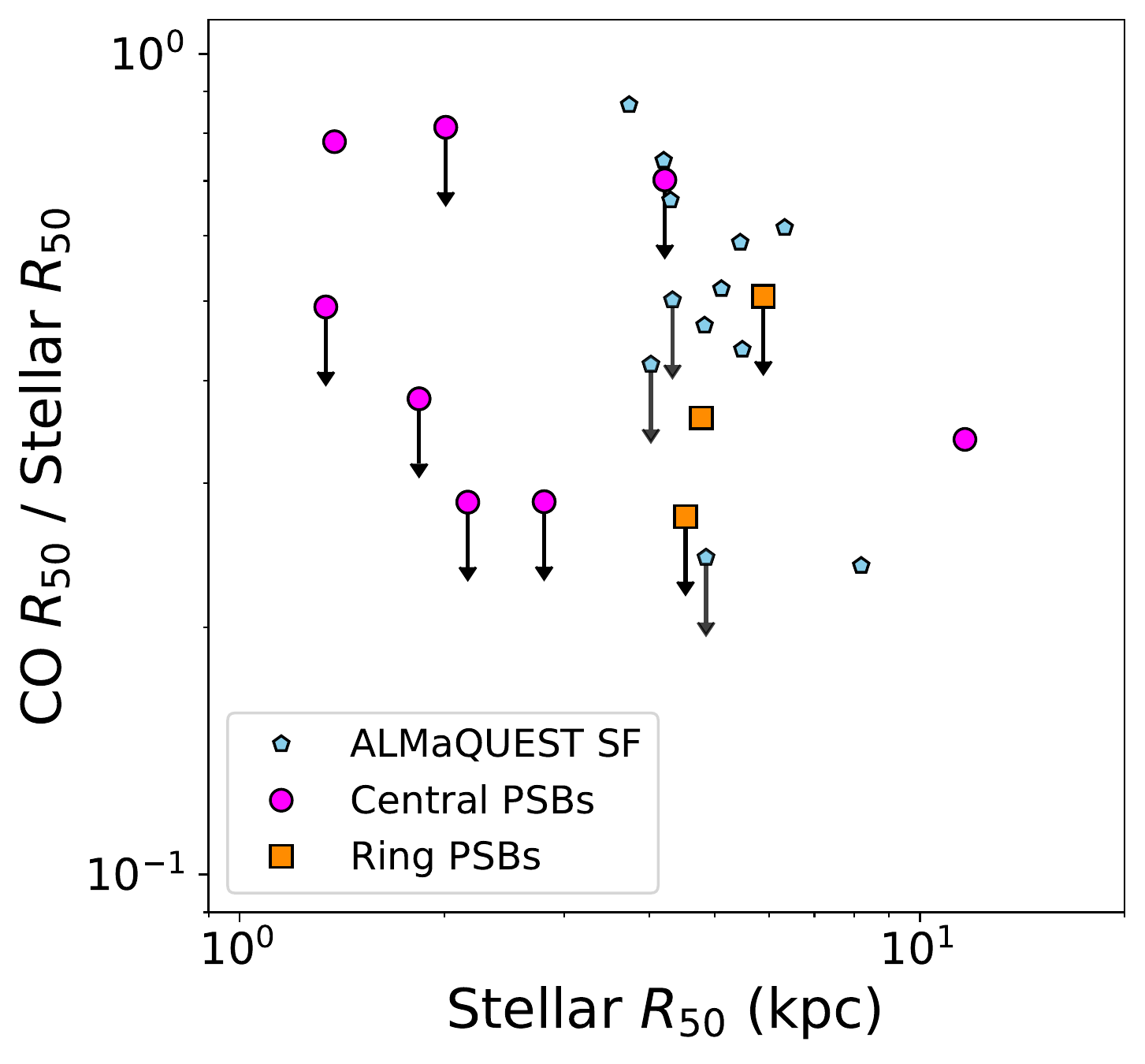}
    \caption{CO half-light radii and stellar $r$-band half-light radii for our PSB sample (centrals in purple, rings in pink) and star-forming galaxies (blue). Galaxies with greater than 50\% of CO emission from the central 3$\sigma$ of the beam are upper limits.}
    \label{fig:r50}
\end{figure}

We use the \texttt{statmorph} python code to measure the half-light radii of the CO emission for our PSBs and the star-forming ALMaQUEST galaxies \citep{rodriguez-gomez_optical_2019}.
We use the half-light radius to measure the overall extent of the CO rather than the central core size measured by the 2D Gaussian fit.
We input a segmentation map of the largest contiguous region of spaxels with 3$\sigma$ CO detections.
We plot the CO and $r$-band half light radii in Figure~\ref{fig:r50}.
Galaxies with $f_{core} \geq 50\%$ are upper limits because at least half of their CO emission is unresolved.
While some of our PSBs cannot be distinguished from the star-forming galaxies due to loosely constraining CO size upper limits, we see that the molecular gas of PSBs is at least as compact as the most compact ALMaQUEST star-forming galaxies, relative to the $r$-band effective radius. 

\section{HI gas properties} \label{app:HI}

\begin{figure*}[hbt!]
    \centering
    \includegraphics[width=0.8\textwidth]{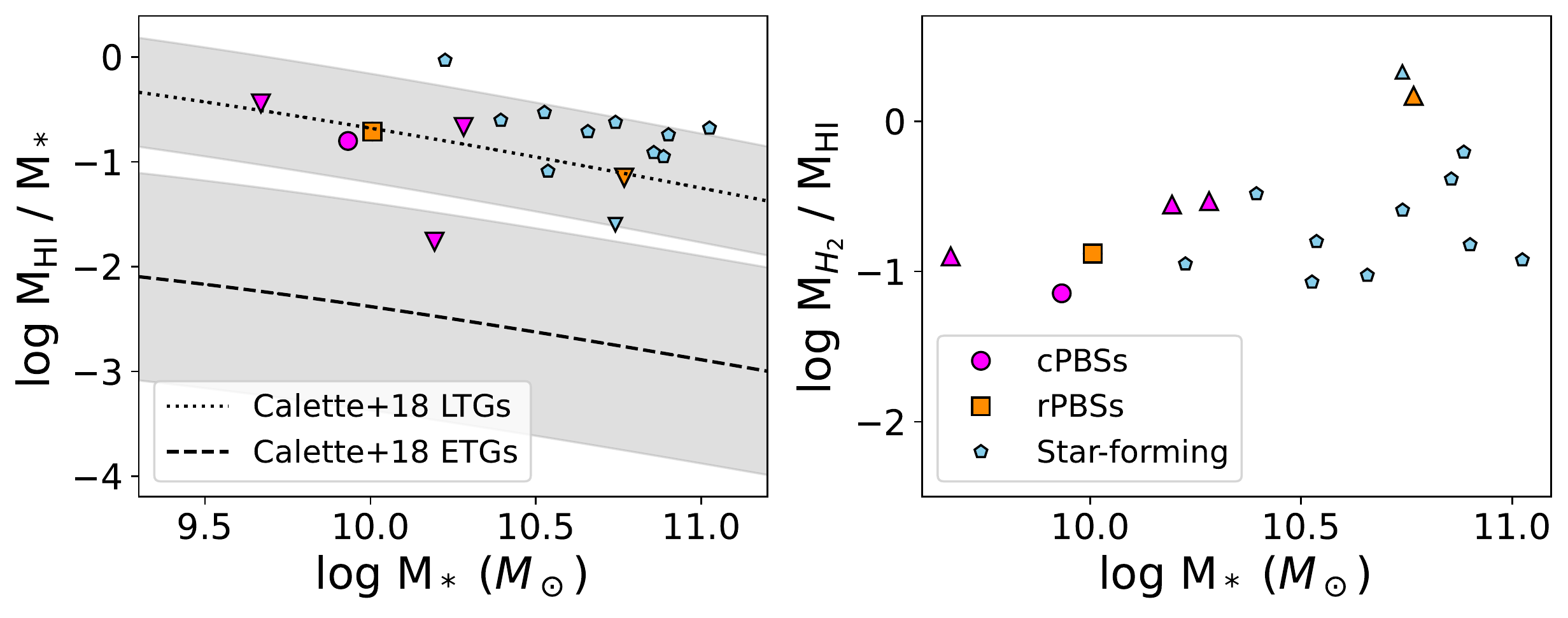}
    \caption{Left: HI gas fraction (M$_{HI}$/M$_*$) versus stellar mass for our cPSBs in purple, rPSBs in orange, and the ALMaQUEST star-forming galaxies in blue. Downward pointing triangles represent M$_{HI}$ upper limits. The dotted and dashed lines show the double power law fits for late-type galaxies and early-type galaxies from \citet{calette_HI_2018} respectively. Right: molecular to atomic gas ratio (M$_{H_2}$/M$_{HI}$) versus stellar mass for the same samples. Upward arrows show M$_{HI}$ upper limits, and downward arrows show M$_{H_2}$ upper limits.}
    \label{fig:hi}
\end{figure*}

We investigate the HI properties of our sample with data from the HI-MaNGA survey \citep{stark_h_2021}, a MaNGA follow-up survey of HI observations with the Green Bank Telescope.
We only include galaxies which have a source confusion probability less than 10\%. 
Of our 13 PSBs, 10 were observed (the other three are outside the redshift cut of $z < 0.05$), and 6/10 were detected.
Only 2/6 detected PSBs have sufficiently low confusion probabilities, 7964-1902 and 8085-6104.
We plot the HI gas fractions (M$_{HI}$/M$_*$) and H$_2$ to HI mass ratio for our PSBs and the ALMaQUEST star-forming galaxies in Figure~\ref{fig:hi}.
We also show the measured HI gas fraction and stellar mass double power law relations for late-type galaxies (LTGs) and early-type galaxies (ETGs) from \citet{calette_HI_2018}.
The two HI-detected PSBs have similar HI gas fractions as the star-forming galaxies, and similar H$_2$ to HI mass ratios.
One PSB, 9494-3701, with an HI mass upper limit has a low HI gas fraction consistent with the ETGs of \citet{calette_HI_2018}.
With only 2 HI detections in our sample, higher resolution HI observations are needed to understand the atomic gas content of our PSBs, especially those with nearby galaxies where source confusion is an issue.

\end{CJK*}
\end{document}